\documentclass[12pt]{article}
\usepackage{amsmath}
\usepackage{amssymb}
\usepackage{epsfig}
\newcommand{\ie}{{i.e.\@ }}
\newcommand{\eg}{{e.g.\@ }}
\newcommand{\Ref}[1]{Ref.\@ \cite{#1}}
\newcommand{\Refs}[1]{Refs.\@ \cite{#1}}
\newcommand{\Eq}[1]{Eq.\@ (\ref{#1})}
\newcommand{\Eqs}[1]{Eqs.\@ (\ref{#1})}
\newcommand{\Sect}[1]{Sect.\@ \ref{#1}}
\newcommand{\Fig}[1]{Fig.\@ \ref{#1}}
\newcommand{\Figs}[1]{Figs.\@ \ref{#1}}
\newcommand{\Tab}[1]{Table \ref{#1}}
\newcommand{\App}[1]{App.\@ \ref{#1}}
\renewcommand{\Im}{\mathrm{Im}\,}
\renewcommand{\Re}{\mathrm{Re}\,}
\newcommand{\tr}{\mathrm{tr}\,}
\newcommand{\MeV}{\;\mathrm{MeV}}
\newcommand{\GeV}{\;\mathrm{GeV}}

\newcommand{\intd}[2]
  {\hspace{-1mm}\int\hspace{-2mm}\frac{d^{#1}{#2}}{(2\pi)^{#1}}\,}
\newcommand{\calD}{{\mathcal D}}
\newcommand{\calL}{{\mathcal L}}
\newcommand{\p}{\pi}
\newcommand{\s}{\sigma}
\renewcommand{\r}{\rho}
\renewcommand{\a}{{a_1}}
\newcommand{\mpi}{m_\pi}
\newcommand{\mpii}[1]{m_{\pi\,{#1}}}
\newcommand{\ms}{m_\sigma}
\newcommand{\mr}{m_\rho}
\newcommand{\ma}{m_{a_1}}
\newcommand{\zp}[1]{Z_{\pi\,{#1}}}
\newcommand{\za}[1]{Z_{a_1\,{#1}}}
\newcommand{\zpa}[1]{Z_{\pi a_1\,{#1}}}
\newcommand{\ovd}[1]{\overline{\delta #1}}
\begin{document}
\begin{titlepage}
\vspace{2.5cm}
\begin{flushright}
May 2001
\end{flushright}
\vspace{3cm}
\begin{center}
{\large \bf Vector and axial-vector correlators in a\\[0.5cm]
  chirally symmetric model}\\[2.5cm]
M. Urban, M. Buballa and J. Wambach\\[0.5cm]
{\small \it Institut f\"ur Kernphysik, TU Darmstadt,
Schlossgartenstr. 9, 64289 Darmstadt, Germany}
\end{center}
\vspace{2.5cm}
\begin{abstract}
  We present a chirally symmetric hadronic model for the vector and
  axial-vector correlators in vacuum. The dominant contributions to
  these correlators come from intermediate pions, $\rho$ and $a_1$
  mesons which are calculated in one-loop approximation. The resulting
  spectral functions are compared with the data obtained by the ALEPH
  collaboration from $\tau$ decay. In the vector channel we find good
  agreement up to $q^2=1\GeV^2$, in the transverse axial-vector
  channel up to $q^2=2\GeV^2$, corresponding to the regimes dominated
  by two-pion or three-pion decay channels, respectively. The
  longitudinal axial correlator is in almost perfect agreement with
  the PCAC result.
\end{abstract}
\end{titlepage}
\setcounter{page}{2}
%
\section{Introduction}
%
The investigation of matter under extreme conditions and the
modifications of hadron properties with density or temperature is one
of the main topics in intermediate and high-energy nuclear
physics. Experimentally, the cleanest information is obtained from
electromagnetic probes which penetrate the medium almost
undisturbed. In the vector dominance model of Sakurai \cite{Sakurai}
electromagnetic processes are mediated by neutral vector mesons
($\rho^0$, $\omega$ and $\phi$). Therefore possible medium
modifications of vector mesons attract particular interest. For
example, the dilepton data measured by the CERES collaboration in
ultra-relativistic nucleus-nucleus collisions \cite{CERES} seem to
give evidence for a change of the $\rho$-meson properties in hot or
dense nuclear matter. The strong enhancement of dilepton production at
low invariant masses ($\sim 0.3 - 0.6\GeV$), was interpreted by Li et
al. \cite{LiKoBrown} as a signature for a dropping $\rho$-meson mass,
conjectured by Brown and Rho \cite{BrownRho} as a consequence of scale
invariance and partial restoration of chiral symmetry. On the other
hand the CERES data could also be explained in an alternative way
by a strong broadening of the $\rho$-meson in the medium due to
resonance formation \cite{FrimanPirner,Peters} and medium
modifications of the pion cloud
\cite{ChanfraySchuck,Herrmann,ChanfrayRappWambach}. (For a review see
\cite{RappWambach}.)

Despite of the phenomenological success of this viewpoint, the role of 
chiral symmetry is not obvious. An important
consequence of chiral symmetry restoration is that the vector and
axial-vector (isovector) correlators, which differ in vacuum due to
spontaneous chiral symmetry breaking, have to become identical in the
restored phase, \ie at high temperature or density. As a precursor,
already at much lower temperatures the correlators start to mix, \ie
the in-medium vector correlator receives contributions proportional to
the vacuum axial-vector correlator and vice versa \cite{DeyEletsky}.

Roughly speaking, the vector correlator is dominated by the $\rho$
meson, while the axial-vector correlator gets important contributions
from the pion and from the $a_1$ meson. In this sense, in the
above-mentioned phenomenological models the medium modification of the
$\rho$-meson could be interpreted as some kind of mixing phenomenon,
caused by thermal or nuclear pions
\cite{ChanfrayDelorme,Krippa}. Certainly, it would be much more
satisfactory, to start from a model which is manifestly chirally
symmetric and to study both, the vector correlator and the
axial-vector correlator simultaneously. Among other things this implies
that we have to include the $a_1$-meson and to treat it on an equal
footing with the $\rho$. Of course, to obtain a realistic description, it
is not sufficient to stay at tree-level, where the $\rho$ and $a_1$
are stable particles without width, but we should also include the
most important decay channels $\rho\rightarrow\pi\pi$ and
$a_1\rightarrow\rho\pi$. This requires at least a one-loop
calculation. The aim of the present paper is to construct such a model
in vacuum, while the application to finite temperatures will be the
topic of a later publication.

In the 1960s the gauged linear $\s$ model emerged as a combination of
the Gell-Mann--L{\'e}vy $\s$ model, which exhibits the essential
features of the spontaneously broken global $SU(2)_L\times SU(2)_R$
chiral symmetry \cite{GellMannLevy}, and Sakurai's idea of vector
meson dominance \cite{Sakurai}, in which a Yang-Mills gauge theory is
constructed with the $SU(2)_V$ isospin group as a gauge group and the
$\r$ meson as the gauge boson. Analogously, Lee and Nieh
\cite{LeeNieh} have built a Lagrangian with local $SU(2)_L\times
SU(2)_R$ symmetry and $\r$ and $\a$ mesons as gauge bosons. The local
(gauge) symmetry is explicitly broken only by the vector meson mass
terms, leading to the current-field identities, and global chiral
symmetry is spontaneously broken as in the ordinary $\s$ model,
leading to the $\r-\a$ mass splitting. In order to render the theory
renormalizable, 't Hooft proposed to generate the vector meson mass by
using the Higgs mechanism rather than by explicitly breaking the local
symmetry \cite{tHooft}.

However, from a modern point of view, there seems to be no reason why
in an effective hadronic model chiral symmetry should be a local
symmetry, while it is only a global symmetry in QCD, the underlying
fundamental theory. In fact, the main advantage of this concept is
that it allows only for a small number of free parameters since all
couplings of gauge fields are determined by only one coupling constant
$g$ \cite{Meissner}. Unfortunately, the ``pure'' gauged linear (or
non-linear) $\s$ model as described above (including the vector-meson
mass term) gives the wrong phenomenology of $\r$ and $\a$ mesons. In
the literature one finds two strategies to cure this problem. The
first one consists of adding terms of higher dimension (\ie with more
fields or derivatives) to the Lagrangian, which are still invariant
under local transformations. One term of this kind was already
considered in \Ref{Gasiorowicz}. In \Ref{Meissner} it was shown that
this is not sufficient, and therefore one more term was added. In a
more recent work \cite{KoRudaz} this strategy was combined with the
second one, which consists of introducing further gauge-invariance
breaking terms in addition to the mass term. Of course, these terms
destroy the current field identity, but we do not believe that this is
a severe objection against this strategy. One could even argue that
the current-field identity cannot be valid in nature, since it is in
contradiction to the large four and six pion continuum observed in the
vector correlator above the $\r$ resonance. In this article we will
completely abandon the concept of local symmetry. We will show that in
this way we can reproduce rather the experimentally measured
spectral functions of the vector and the axial-vector correlators
without adding terms of higher dimension.

The paper is organized as follows. In \Sect{SectL} we introduce the
Lagrangian of our model. We begin with a brief review of the
properties of the gauged linear $\s$ model in
\Sect{SectLgauged}. Then, in \Sect{SectLglobal}, we relax the
constraints imposed by the gauge symmetry, retaining only a global
chiral symmetry. Our model is complete after including electroweak
interactions in \Sect{SectLelectroweak}. In \Sect{Sectoneloop} we
construct the propagators of the mesons ($\pi$, $\s$,
$\rho$ and $a_1$) in a one-loop approximation. Here special care is
taken, not to destroy the symmetries of the model by an unsuited
regularization scheme. Since we give up the current-field identity,
the vector and axial-vector currents are not just proportional to the
fields, but involve additional terms. The corresponding
correlation functions are derived in \Sect{Sectcorrelators}. In
\Sect{Sectresults} we present numerical results for the vector and
axial-vector correlator, and related observables, such as the
electromagnetic form factor of the pion and the weak $\tau$ decay. Our
results are compared with experimental data. Finally, in
\Sect{Sectconclusions}, we draw conclusions.
%
\section{Linear \boldmath{$\s$} models with vector mesons}
%
\label{SectL}
In this section we want to construct the chirally symmetric Lagrangian 
which will be the starting point for our analysis. In view of the 
phenomenological success of the vector dominance model \cite{Sakurai} 
in the vector channel, the gauged linear $\s$ model \cite{LeeNieh}, 
which is its generalization to $SU(2)_L\times SU(2)_R$ chiral symmetry, 
seems to be a natural candidate. In the literature, this model served
\eg as a toy model for the calculation of $\rho$ and $a_1$ meson
properties in a hot meson gas \cite{Pisarski}. However, as we will discuss,
the gauged linear $\s$ model does not correctly reproduce the phenomenology 
of $\rho$ and $a_1$ mesons in vacuum, and we will have to modify it. It is 
nevertheless useful to begin with a brief review of the basics of this model
because most of the definitions and formulas given here will remain valid 
after our modifications.
%
\subsection{The gauged linear \boldmath{$\s$} model}
\label{SectLgauged}
Defining the four-component field
\begin{equation}
\Phi=\begin{pmatrix}\s\\ \vec{\p}\end{pmatrix}\, ,
\end{equation}
we can write the Lagrangian of the mesonic part of the linear $\s$
model as
\begin{equation}
\calL_\Phi = \frac{1}{2}\,\partial_\mu\Phi\cdot\partial^\mu\Phi
  -\frac{\mu^2}{2}\,\Phi\cdot\Phi
  -\frac{\lambda^2}{4}\,(\Phi\cdot\Phi)^2\, .
\label{L0}
\end{equation}
For $\mu^2 < 0$ chiral symmetry is spontaneously broken, yielding
massless pions. To obtain massive pions, we add a small explicit
symmetry breaking term:
\begin{equation}
\calL_\mathrm{SB} = c \s\, .
\end{equation}
Let $\vec{T}$ and $\vec{T}^5$ be $4\times 4$ matrices acting in the
$(\s,\vec{\p})$ space, with $T_i$ generating ordinary isospin rotation
around the $\p_i$ axis and $T_i^5$ inducing rotations in the $\s-\p_i$
plane. They fulfill the commutation relations
\begin{equation}
[T_i\,,T_j] = i \varepsilon_{ijk}\, T_k\, ,\qquad
[T_i\,,T_j^5] = i \varepsilon_{ijk}\, T_k^5\, ,\qquad
[T_i^5\,,T_j^5] = i \varepsilon_{ijk}\, T_k\, ,
\label{TCommutators}
\end{equation}
and are normalized to
\begin{equation}
\tr T_i T_j = 2\delta_{ij}\, ,\qquad
\tr T_i T_j^5 = 0\, ,\qquad
\tr T_i^5 T_j^5 = 2\delta_{ij}\, .
\label{TNormalization}
\end{equation}
One can easily see that these operators generate an $SU(2)_L\times
SU(2)_R$ group by inspecting the commutation relations of the
combinations
\begin{equation}
T^L_i = \frac{1}{2}\,(T_i-T^5_i)\, ,\qquad
T^R_i = \frac{1}{2}\,(T_i+T^5_i)\, .
\end{equation}
 
The Lagrangian (\ref{L0}) is invariant under global $SU(2)_L\times
SU(2)_R$ transformations $U$,
\begin{equation}
\Phi \rightarrow U\Phi\, ,\qquad
U = e^{i\vec{\alpha}\cdot\vec{T} + i\vec{\beta}\cdot\vec{T}^5}\, .
\end{equation}
To make it invariant also under local transformations $U(x)$, we
introduce gauge fields $Y_\mu$\,,
\begin{equation}
Y_{\mu} = \vec{\r}_\mu\cdot \vec{T} + \vec{a}_{1\,\mu}\cdot\vec{T}^5\, ,
\end{equation}
and replace the ordinary derivatives in \Eq{L0} by ``covariant
derivatives'',
\begin{equation}
\calD_\mu \Phi = (\partial_\mu - i g\,Y_\mu) \Phi = \begin{pmatrix}
  \partial_\mu\s + g\,\vec{a}_{1\,\mu}\cdot\vec{\p}\\
  \partial_\mu\vec{\p} + g\,\vec{\r}_\mu\times\vec{\p}
    -g\,\vec{a}_{1\,\mu}\s
\end{pmatrix}\, .
\end{equation}
$\calD_\mu \Phi$ behaves under local transformations in the same way
as the field $\Phi$, provided the gauge fields $Y_\mu$ transform as
\begin{equation}
Y_\mu\rightarrow U Y_\mu U^\dagger+\frac{i}{g}\,U\partial_\mu U^\dagger\, .
\label{Ytransform}
\end{equation}

The field strength tensor of the gauge fields is defined in the usual way as
\begin{equation}
Y_{\mu\nu} = \partial_\mu Y_\nu - \partial_\nu Y_\mu + i
  g\,[Y_\mu\,,Y_\nu] =: \vec{\r}_{\mu\nu}\cdot \vec{T} +
  \vec{a}_{1\,\mu\nu}\cdot\vec{T}^5\, ,
\end{equation}
with
\begin{align}
\vec{\r}_{\mu\nu}&=\partial_\mu\vec{\r}_\nu-\partial_\nu\vec{\r}_\mu
  +g\,\vec{\r}_\mu\times\vec{\r}_\nu
  +g\,\vec{a}_{1\,\mu}\times\vec{a}_{1\,\nu}\, ,
\nonumber\\
\vec{a}_{1\,\mu\nu}&=\partial_\mu\vec{a}_{1\,\nu}-\partial_\nu\vec{a}_{1\,\mu}
  +g\,\vec{a}_{1\,\mu}\times\vec{\r}_\nu
  +g\,\vec{\r}_\mu\times\vec{a}_{1\,\nu}\, ,
\end{align}
and transforms as $Y_{\mu\nu}\rightarrow U Y_{\mu\nu}
U^\dagger$. Therefore the Yang-Mills term
\begin{equation}
\calL_{Y} = -\frac{1}{8}\, \tr Y_{\mu\nu}Y^{\mu\nu} =
  -\frac{1}{4}\,(\vec{\r}_{\mu\nu}\cdot\vec{\r}^{\,\mu\nu}
  +\vec{a}_{1\,\mu\nu}\cdot\vec{a}_1^{\,\mu\nu})
\label{LkinY0}
\end{equation}
is obviously gauge invariant.

Finally we add a mass term for the vector mesons, which is invariant
under global transformations, but breaks gauge invariance:
\begin{equation}
\calL_{m_0} = \frac{m_0^2}{4}\, \tr Y_\mu Y^\mu
  = \frac{m_0^2}{2}\,(\vec{\r}_\mu \cdot \vec{\r}^{\,\mu}
  +\vec{a}_{1\,\mu} \cdot \vec{a}_1^{\,\mu})
\end{equation}

Combining everything, we obtain the final Lagrangian:
\begin{equation}
\calL_\mathrm{gauged}
=\, (\calL_\Phi)_{\partial_\mu\Phi \rightarrow 
  \calD_\mu\Phi}+\calL_\mathrm{SB}+\calL_{Y}+\calL_{m_0}
\label{Lgauged}
\end{equation}

As a next step we will briefly describe how this Lagrangian is usually
treated. Due to spontaneous symmetry breaking the $\s$ field has a
non-vanishing expectation value, so that it is convenient to redefine
the $\s$ field,
\begin{equation}
\s_\mathrm{new} = \s_\mathrm{old}-\s_0\, ,\qquad
\Phi_\mathrm{new} = \Phi_\mathrm{old}-
  \begin{pmatrix}\s_0\\ \vec{0}\end{pmatrix}
  = \Phi_\mathrm{old}-\Phi_0\, ,
\label{shiftsigma}
\end{equation}
and to adjust $\s_0$ such that the expectation value of the new $\s$
field vanishes (\ie $\s_0$ is the expectation value of the old $\s$
field). At tree level, this is achieved by minimizing the potential,
\ie by solving the equation
\begin{equation}
\lambda^2 \s_0^3+\mu^2 \s_0-c=0\, .
\label{minimumsigma0}
\end{equation}
After the shift of the $\s$ field the masses of pion and $\s$ and the
masses of $\r$ and $\a$ are no longer degenerate:
\begin{equation}
\mpii{0}^2 = \mu^2+\lambda^2 \s_0^2\, ,\quad
\ms^2 = \mu^2+3 \lambda^2 \s_0^2\, ,\quad
\mr^2 = m_0^2\, ,\quad
\ma^2 = m_0^2+g^2 \s_0^2\, .
\label{masses0}
\end{equation}
\Eq{minimumsigma0} can now be written as $\mpii{0}^2\,\s_0-c=0$, \ie
for $c=0$ the pions are massless, as required by the Goldstone
theorem.

In addition to the mass splitting, the shift of the $\s$ field
generates many new vertices, in particular a $\p-\a$ vertex generated
by the interaction Lagrangian
\begin{equation}
\calL_{\p\a} = -i g\,Y_\mu\Phi_0\cdot\partial^\mu\Phi
  = -g \s_0\,\vec{a}_{1\,\mu}\cdot\partial^\mu\vec{\p}\, .
\label{Lpia1mix}
\end{equation}
This leads to the so-called $\p-\a$ mixing, which is usually
eliminated by a redefinition of the $\a$ field,
$\vec{a}_{1\,\mu\,\mathrm{new}} = \vec{a}_{1\,\mu\, \mathrm{old}} + g
\s_0/\ma^2 \partial_\mu \vec{\p}$\,, followed by a wave-function
renormalization of the pion field, $\vec{\p}_\mathrm{new} =
\vec{\p}_\mathrm{old}/\sqrt{Z}$, with $Z = \ma^2/\mr^2$\,. As a
consequence, the physical pion mass is increased by $\sqrt{Z}$,
\begin{equation}
\mpi = \frac{\ma}{\mr}\,\mpii{0}\, ,
\label{mpi0}
\end{equation}
whereas the pion decay constant is reduced by $1/\sqrt{Z}$,
\begin{equation}
f_\p = \frac{\mr}{\ma}\,\s_0\, .
\label{fpi0}
\end{equation}
However, we will choose a different method to treat the $\p-\a$
mixing, namely summing the self-energy diagrams generated by the
$\p-\a$ vertex to all orders, which gives the same results at tree
level, but is more convenient if loop corrections are taken into
account. In the next section this will be discussed in more detail.

\begin{figure}
\begin{center}
\epsfig{file=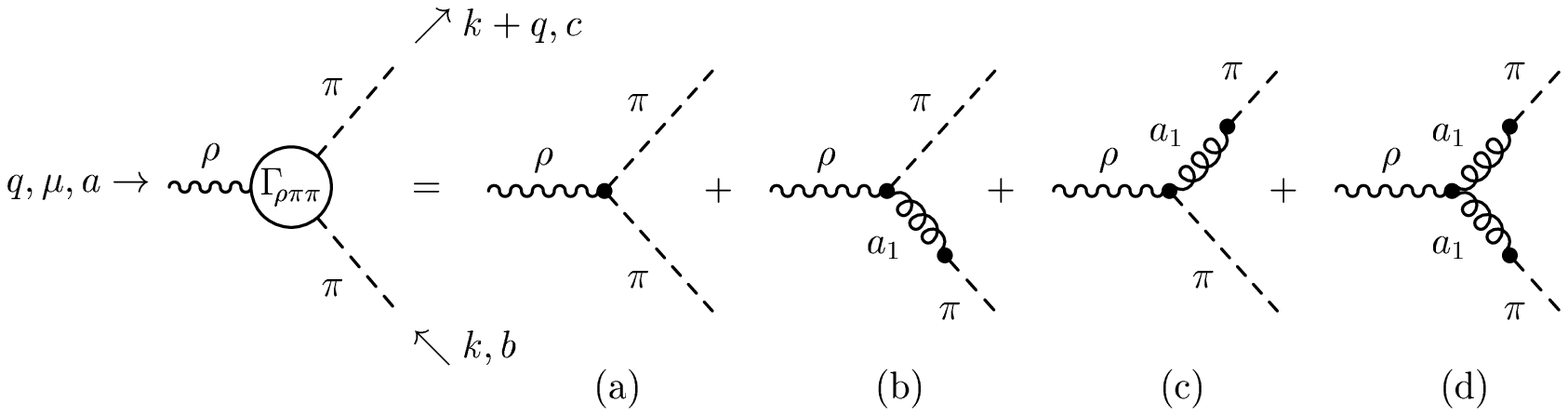,width=13.8cm,bbllx=48,bblly=649,bburx=532,bbury=776}
\end{center}
\vspace{-.5cm}
\caption{\small The $\r\p\p$ vertex function of the gauged linear $\s$ model
  after symmetry breaking.}
\label{FigGrpp}
\end{figure}

\begin{figure}
\begin{center}
\epsfig{file=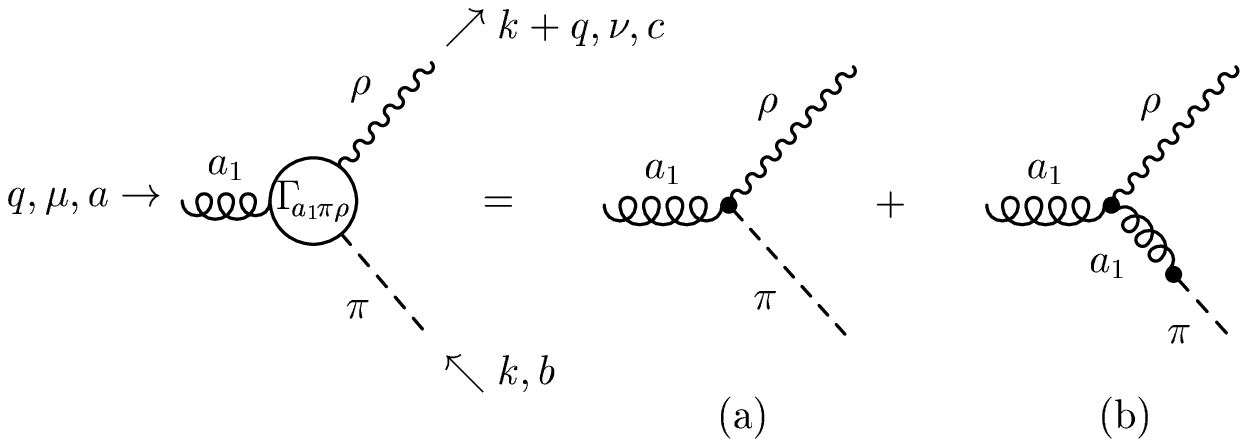,width=10.2cm,bbllx=63,bblly=649,bburx=421,bbury=776}
\end{center}
\vspace{-.5cm}
\caption{\small The $\a\p\r$ vertex function of the gauged linear $\s$ model 
  after symmetry breaking.}
\label{FigGapr}
\end{figure}

The properties of the $\r$ and the $\a$ are dominated by the decay modes
$\r\rightarrow \p\p$ and $\a\rightarrow \p\r$, respectively. The
corresponding vertex functions within the present model are shown in
\Figs{FigGrpp} and \ref{FigGapr}.
If the $\p-\a$ mixing is not eliminated, there are four
contributions to the $\r\p\p$ vertex. The corresponding vertex
factors read:
\begin{align}
\big(\Gamma_{\r\p\p}^\mathrm{(a)}\big)_{abc}^{\mu}
  &=-g\,\varepsilon_{abc}\,(2\,k^\mu+q^\mu)\, ,
\label{Grppa}\\*
\big(\Gamma_{\r\p\p}^\mathrm{(b+c)}\big)_{abc}^{\mu}
  &=\frac{g^3 \s_0^2}{\ma^2}\,\varepsilon_{abc}\,(2\,k^\mu+q^\mu)\, ,
\label{Grppbc}\\*
\big(\Gamma_{\r\p\p}^\mathrm{(d)}\big)_{abc}^{\mu}
  &=\frac{g^3 \s_0^2}{\ma^4}\,\varepsilon_{abc}\,
  (q^2\,k^\mu-k\cdot q\,q^\mu)\, .
\label{Grppd}
\end{align}
The coupling constant $g$ can be determined from \Eqs{masses0} and
(\ref{fpi0}) and the empirical values of $f_\pi$, $\mr$ and $\ma$. If
we take \eg $\mr=770\MeV$, $\ma\approx\sqrt{2}\,\mr=1090\MeV$\,, and
$f_\p = 93\MeV$, we find $g=5.9$\,. This is more or less the value
which is needed to obtain a good description of the $\rho$-meson width
and the electromagnetic pion form factor in models without
$a_1$-degrees of freedom, where the $\r\p\p$ vertex function consists
of diagram (a) only (see \eg \Ref{Herrmann}). In the present model,
the second contribution, (b+c), has the same structure as (a) 
and leads to a reduction of the $\r\p\p$ coupling constant
by a factor $1-g^2 \s_0^2/\ma^2=\mr^2/\ma^2=1/Z$. However, this is
exactly cancelled by the factors $\sqrt{Z}$ assigned to each external
pion line if one calculates the decay $\r\rightarrow\p\p$. Thus, the
contributions (a) to (c) would give a quite good description of the
decay $\r\rightarrow\p\p$ with the value $g=5.9$ determined above.
Unfortunately there is still contribution (d), which generates a
strong $q^2$-dependence of the vertex function. This does not only
result in a too small width for $\r\rightarrow\p\p$, but also leads to
a wrong shape of the pion electromagnetic form factor or the $\p-\p$
scattering phase shifts in the $J=1, I=1$ channel as a function of
$q^2$, which can perfectly be reproduced without momentum dependence
of the vertex.

A similar effect exists for the decay $\a\rightarrow\p\r$. Due to
$\p-\a$ mixing there are two contributions to the $\a\p\r$ vertex
function shown in \Fig{FigGapr}:
\begin{align}
\big(\Gamma_{\a\p\r}^{(a)}\big)_{abc}^{\mu\nu}
  &=i g^2\s_0\,\varepsilon_{abc}\,g^{\mu\nu}
 \, ,\\
\big(\Gamma_{\a\p\r}^{(b)}\big)_{abc}^{\mu\nu}
  &=-i \frac{g^2 \s_0}{\ma^2}\,\varepsilon_{abc}\,
  \big(k^\mu k^\nu+k^\mu q^\nu+q^\mu k^\nu-(k^2+2\,k\cdot q)\,g^{\mu\nu}\big)
  \, .
\end{align}
Again, the term arising from the non-abelian terms in the field
strength tensors $Y_{\mu\nu}$,
\begin{equation}
\calL_{YYY}=-\frac{i g}{2}\,\tr \partial_\mu Y_\nu\,[Y^\mu,Y^\nu]
\end{equation}
causes an additional momentum dependence of the vertex. As a
consequence, the amplitude for $\a\rightarrow\r\p$ as a function of
the invariant mass of the $\a$ has a zero at $q^2 =
\ma^2+\mr^2$\,. This lies well below the $\tau$ mass, \ie within the
range where the $\a$ can experimentally be observed under clean
conditions. The experimental data, however, show no minimum in the
$\p\r$ invariant mass spectrum \cite{ARGUS,ALEPH}.
%
\subsection{Restriction to global chiral symmetry}
%
\label{SectLglobal}
The problems discussed in the end of the previous section force us to
modify the Lagrangian, \Eq{Lgauged}. To that end we relax the
constraints imposed by the requirement that the Lagrangian (except for
the mass term ${\cal L}_{m_0}$) should be invariant under a local
$SU(2)_L\times SU(2)_R$ transformation and only insist on a global
chiral symmetry. In fact, this has already been done in \Ref{KoRudaz},
but in a less radical way.

Two consequences of the gauge principle are incorporated in the
Lagrangian (\ref{Lgauged}): First, the same coupling constant $g$
appears in different interaction terms, and second, some chirally
symmetric interaction terms are not allowed. For example, the most
general interaction of two $\Phi$ fields with two $Y$ fields which is
chirally symmetric and has dimension $\leq 4$ looks as follows:
\begin{equation}
\calL_{\Phi\Phi YY} = -\frac{h_1}{2}\,Y_\mu \Phi\cdot Y^\mu \Phi
  +\frac{h_2}{4}\,\Phi\cdot\Phi\,\tr Y_\mu Y^\mu\, .
\end{equation}
The gauge principle would force us to choose $h_1 = g^2$ and $h_2 =
0$, whereas in \Ref{KoRudaz}, $h_1$ and $h_2$ were chosen as
independent parameters $h_1 = g^2\,(1+2 c)$ and $h_2 = g^2\,(b-c)$
(with $b$ and $c$ as defined in \Ref{KoRudaz}).

As we have shown in the previous section, the problems with the
phenomenology of $\r$ and $\a$ mesons arise mainly from the vector
meson self interaction. However, if we are interested only in global
chiral symmetry, we can adjust the coupling constants independently:
\begin{equation}
\calL_{YYY}+\calL_{YYYY} = 
  -\frac{i g^\prime}{2}\,\tr \partial_\mu Y_\nu\,[Y^\mu,Y^\nu]
  +\frac{g^{\prime\prime\,2}}{4}\,\tr Y_\mu Y_\nu\,[Y^\mu,Y^\nu]\, .
\label{LYYYLYYYY}
\end{equation}
We are allowed to choose the coupling constants $g^\prime$ and
$g^{\prime\prime}$ smaller than $g$ or even to put them to zero. In
fact, we perform our calculations with arbitrary values of these
coupling constants, and find that the best description of the
axial-vector correlator can be achieved with $g^\prime=0$, while our
results are insensitive to $g^{\prime\prime}$. For simplicity, we will
put both, $g^\prime$ and $g^{\prime\prime}$, to zero throughout this
paper.

In addition to the $g^{\prime\prime}$ term, one could think of other
chirally symmetric $YYYY$-interaction terms of dimension $4$. However,
they are not important for our purposes, either. Higher-dimension
terms, which are frequently used in the literature
\cite{Gasiorowicz,Meissner,KoRudaz}, are much more complicated and
result in momentum dependent vertices. As we will see, the
experimental data on $\rho$ and $a_1$ mesons can very well be
explained without such terms. This indicates that in nature the
momentum dependence of the vertices is small.

Besides the interaction terms, we also modify the kinetic term for the
vector mesons contained in \Eq{LkinY0}. With the original kinetic
term, the longitudinal part of the vector meson propagator does not go
to zero for $k^2\rightarrow \infty$, causing serious problems if one
tries to calculate loop diagrams with vector meson lines. To avoid
these problems, we take the St{\"u}ckelberg Lagrangian
\cite{ItzyksonZuber}
\begin{equation}
\calL_{\mathrm{kin}\,Y} = -\frac{1}{8}\,\tr(\partial_\mu
  Y_\nu-\partial_\nu Y_\mu) (\partial^\mu Y^\nu-\partial^\nu Y^\mu)
  -\frac{\xi}{4}\,\tr(\partial_\mu Y^\mu)^2\, .
\end{equation}
From this one retrieves the kinetic term of the original Lagrangian
(\ref{LkinY0}) by taking the limit $\xi\rightarrow 0$\,. The $\r$
meson propagator now has the form
\begin{equation}
G^{\mu\nu}_\r(k) =
  \frac{\frac{k^\mu k^\nu}{\mr^2}-g^{\mu\nu}}{k^2-\mr^2}
  -\frac{\frac{k^\mu k^\nu}{\mr^2}}{k^2-\mr^2/\xi}\, .
\end{equation}
This form has the advantage that for $\xi > 0$ the propagator behaves
as $1/k^2$ for $k^2\rightarrow \infty$, which is not true for $\xi =
0$. On the other hand, we have introduced unphysical scalar particles
with mass $\mr/\sqrt{\xi}$. This can be interpreted as some form of
regularization of the contributions from the longitudinal polarization
states of the vector mesons. Therefore the parameter $\xi$ must be
chosen small enough, such that all thresholds involving these
unphysical particles lie above the energy range we are interested in%
\footnote{In a true gauge theory the final results are independent of
$\xi$, because in this case the contributions of the unphysical scalar
particles are cancelled by the ghost contributions.}%
.

Combining everything, we obtain our full Lagrangian:
\begin{align}
\calL=\,&\frac{1}{2}\,\partial_\mu\Phi\cdot\partial^\mu\Phi
  -\frac{\mu^2}{2}\,\Phi\cdot\Phi 
  -i g\, Y_\mu \Phi\cdot\partial^\mu\Phi
  -\frac{h_1}{2}\,Y_\mu \Phi\cdot Y^\mu \Phi
  +\frac{h_2}{4}\,\Phi\cdot\Phi\,\tr Y_\mu Y^\mu
\nonumber\\ 
& -\frac{\lambda^2}{4}\,(\Phi\cdot\Phi)^2+c \s
  -\frac{1}{8}\,\tr(\partial_\mu Y_\nu-\partial_\nu Y_\mu)
    (\partial^\mu Y^\nu-\partial^\nu Y^\mu)
  -\frac{\xi}{4}\,\tr(\partial_\mu Y^\mu)^2
\nonumber\\
& +\frac{m_0^2}{4}\,\tr Y_\mu Y^\mu\, .
\label{Lglobal}
\end{align}

Again, as in the previous section, we must shift the $\s$ field, and
\Eqs{shiftsigma} and (\ref{minimumsigma0}) remain valid. But instead
of \Eq{masses0} we find the following expressions for the meson
masses:
\begin{equation}
\mpii{0}^2 = \mu^2+\lambda^2 \s_0^2\, ,\quad
\ms^2 = \mu^2+3 \lambda^2 \s_0^2\, ,\quad
\mr^2 = m_0^2+h_2\,\s_0^2\, ,\quad
\ma^2 = m_0^2+(h_1+h_2)\,\s_0^2\, .
\label{masses}
\end{equation}
This has the interesting consequence that with a special choice of
parameters ($m_0 = 0$, $h_2 = \mr^2/\s_0^2$) it is now possible to
generate the vector meson masses from chiral symmetry breaking alone.
In this case the masses of both $\r$ and $\a$ meson would drop in a
nuclear medium where chiral symmetry is (partially) restored
(``Brown--Rho scaling'' \cite{BrownRho}).

\begin{figure}
\begin{center}
\epsfig{file=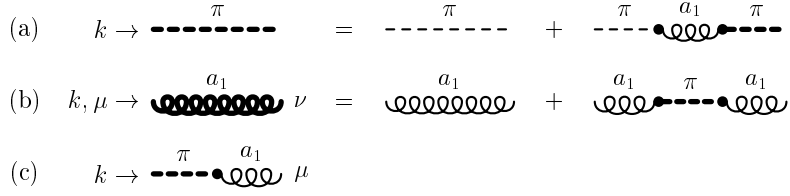,width=10.4cm,bbllx=149,bblly=655,bburx=515,bbury=744}
\end{center}
\vspace{-0.5cm}
\caption{\small Propagators with $\p-\a$ mixing: (a) the pion propagator
  $i G_\p(k)$, (b) the $\a$ propagator $i G_{\a}^{\mu\nu}(k)$, and (c) 
  the mixed propagator $i G_{\p\a}^{\mu}(k)$.}
\label{Figpia1mix}
\end{figure}

The $\p-\a$ mixing Lagrangian $\calL_{\p\a}$ is still given by
\Eq{Lpia1mix}, but because of the St{\"u}ckelberg ($\xi$-) term in the
Lagrangian it cannot be eliminated by a simple field
redefinition. Instead, self-energy contributions from $\calL_{\p\a}$
must be iterated to all orders as shown in \Fig{Figpia1mix}(a).
Then the pion propagator can be written as
\begin{equation}
G_\p(k)
  = \frac{1}{k^2-\mpii{0}^2
  +\frac{g^2 \s_0^2 k^2}{\xi k^2-\ma^2}}
  = \sum_{i=1}^2 \frac{\zp{i}}{k^2-\mpii{i}^2}\, .
\end{equation}
The first pole at $\mpii{1}=\mpi$ corresponds to the physical pion,
whereas the second one at $\mpii{2}$ corresponds to the unphysical
longitudinally polarized $\a$ particle. In principle $\mpii{1}$ and
$\mpii{2}$ are functions of $\mpii{0}$\,, but it is more convenient to
quote the relations for these masses as functions of the physical pion
mass:
\begin{equation}
\mpii{1}^2=\mpi^2\, ,\quad
\mpii{2}^2=\frac{\ma^2 (\ma^2-g^2 \s_0^2-\xi\mpi^2)}
  {\xi (\ma^2-\xi\mpi^2)}\, ,\quad
\mpii{0}^2=\frac{\mpi^2 (\ma^2-g^2 \s_0^2-\xi\mpi^2)}
  {\ma^2-\xi\mpi^2}\, .
\label{mpi012}
\end{equation}
The $\zp{i}$ are given by
\begin{equation}
\zp{1}=
  \frac{\ma^2-\xi \mpii{1}^2}{\xi (\mpii{2}^2-\mpii{1}^2)}\, ,\qquad
\zp{2}=1-\zp{1}\, .
\end{equation}
Here $\zp{1}$ corresponds to the $Z$ factor introduced in the previous
section. In the limit $\xi\rightarrow 0$, $h_1\rightarrow g^2$ and
$h_2\rightarrow 0$ we retrieve the relations for $\mpi$, $\mpii{0}$
and $Z$ in the gauged linear sigma model.

The $\a$ and the mixed ($\p-\a$) propagator can be obtained from the
pion propagator as shown in \Figs{Figpia1mix}(b) and (c). For the $\a$
propagator we find
\begin{equation}
G_{\a}^{\mu\nu}(k)=
  \frac{\frac{k^\mu k^\nu}{\ma^2}-g^{\mu\nu}}{k^2-\ma^2}
  +k^\mu k^\nu \sum_{i=1}^2\frac{\za{i}}{k^2-\mpii{i}^2}\, ,
\end{equation}
with
\begin{equation}
\za{1}=\frac{\mpii{1}^2-\mpii{0}^2}
  {\xi \mpii{1}^2 (\mpii{2}^2-\mpii{1}^2)}\, ,\qquad
\za{2}=-\frac{1}{\ma^2}-\za{1}\, ,
\end{equation}
and for the mixed propagator
\begin{equation}
G_{\p\a}^\mu(k)=
  i k^\mu \sum_{i=1}^2\frac{\zpa{i}}{k^2-\mpii{i}^2}\, ,
\end{equation}
with 
\begin{equation}
\zpa{1}=-\frac{g \s_0}{\xi (\mpii{2}^2-\mpii{1}^2)}\, ,\qquad
\zpa{2}=-\zpa{1}\, .
\end{equation}

Since we have switched off the 3-vector interaction,
for the $\r\p\p$ vertex functions only the diagrams (a) to (c) of
\Fig{FigGrpp} are left. If the external pions are on-shell, \ie
$k^2=(k+q)^2=\mpi^2$\,, the total vertex function reads,
\begin{equation}
\big(\Gamma_{\r\p\p}\big)_{abc}^{\mu}
  =-\Big(g-\frac{h_1\s_0^2}{\ma^2-\xi\mpi^2}\Big)\,\varepsilon_{abc}\,
    (2\,k^\mu+q^\mu)\, .
\end{equation}
This can be used to define an effective $\r\p\p$ coupling constant
\begin{equation}
g_{\r\p\p}^\mathrm{eff} 
  = \Big(g-\frac{h_1\s_0^2}{\ma^2-\xi\mpi^2}\Big) \zp{1}
  = g \frac{(\mr^2-\xi\mpi^2)(\ma^2-\xi\mpi^2)}
    {(\ma^2-\xi\mpi^2)^2-g^2 \s_0^2 \ma^2}\, .
\label{grhoeff}
\end{equation}
%
\subsection{Electroweak interactions}
%
\label{SectLelectroweak}
Until now we have completely neglected the electroweak
interaction. However, the cleanest experimental information about $\r$
and $\a$ mesons can be obtained from electromagnetic reactions,
$e^+e^-\rightarrow\r$, and from the weak decay of the $\tau$ lepton,
$\tau\rightarrow\nu_\tau\,\r$ or
$\tau\rightarrow\nu_\tau\,\a$. Another reason why we want to include
the electromagnetic interactions is of course our interest in dilepton
production in heavy ion collisions.

At the quark level there is no ambiguity how the electromagnetic field
and the $W$ bosons must be introduced: The ordinary derivative of the
quark fields is replaced by a covariant derivative. If we consider
only $u$ and $d$ quarks ($\psi = (u, d)$), this looks as follows (see
\eg \cite{PDB}):
\begin{equation}
D_{\mu}\psi
  =\Big(\partial_\mu-i e\,A_\mu\,\frac{\tau_3}{2}
  -\frac{i e\cos\theta_C}{\sin\theta_W}\,\frac{1-\gamma_5}{2}\,
  \Big(W_{1\,\mu}\frac{\tau_1}{2}+W_{2\,\mu}\frac{\tau_2}{2}\Big)
  +\dots\Big)\psi\,\, .
\end{equation}
Here we have omitted the isoscalar part of the electromagnetic
interaction and the $Z$ boson coupling, since they are irrelevant for
our purposes.

This can be generalized to the $(\s,\vec{\p})$ meson sector. The
isospin operator $\vec{\tau}/2$ corresponds to the $\vec{T}$ operator,
whereas $\gamma_5\vec{\tau}/2$ can be identified with
$\vec{T}^5$. Therefore all derivatives $\partial_{\mu}\Phi$ in the
Lagrangian (\ref{Lglobal}) have to be replaced by
\begin{align}
D_\mu\Phi&=\Big(\partial_\mu-i e\,A_\mu T_3
  -\frac{i e \cos\theta_C}{\sin\theta_W}\,
  (W_{1\,\mu}T^L_1+W_{2\,\mu}T^L_2)\Big)\Phi
\nonumber\\
& =\begin{pmatrix}\partial_\mu\,\s\\ \partial_\mu\,\vec{\p}\end{pmatrix}
  +e\begin{pmatrix}0\\ \vec{A}_\mu\times\vec{\p}\end{pmatrix}
  +\frac{e\cos\theta_C}{2\sin\theta_W}
  \begin{pmatrix}-\vec{W}_\mu\cdot\vec{\p}\\
  \vec{W_\mu}\times\vec{\p}+\vec{W}_\mu\,\s\end{pmatrix}\, .
\label{DPhi}
\end{align}
In the last line we have used the abbreviations
\begin{equation}
\vec{A}_\mu = \begin{pmatrix}0\\ 0\\ A_\mu\end{pmatrix}\, ,\qquad
\vec{W}_\mu = \begin{pmatrix}W_{1\,\mu}\\ W_{2\,\mu}\\ 0\end{pmatrix}\, .
\end{equation}
To find the operators in the vector-meson space corresponding to the
$\vec{T}$ and $\vec{T}^5$ operators in the $(\s,\vec{\p})$-meson
space, we can expand \Eq{Ytransform} for the case $\partial_\mu U=0$
to linear order in $\vec{\alpha}$ and $\vec{\beta}$. Then it becomes
clear that the derivatives of the vector meson fields, $\partial_\mu
Y_\nu$\,, must be changed into
\begin{align}
D_\mu Y_\nu&=\partial_\mu Y_\nu-i e\,A_\mu\,\lbrack T_3\,,Y_\nu\rbrack
  -\frac{i e \cos\theta_C}{\sin\theta_W}\,
  (W_{1\,\mu}\,\lbrack T^L_1\,,Y_\nu\rbrack
  +W_{2\,\mu}\,\lbrack T^L_2\,,Y_\nu\rbrack)
\nonumber\\
& =:D_\mu\vec{\r}_\nu\cdot\vec{T}+D_\mu\vec{a}_{1\,\nu}\cdot\vec{T}^5\, ,
\label{DY}
\end{align}
with
\begin{align}
D_\mu\vec{\r}_\nu&=
  \partial_\mu\,\vec{\r}_\nu+e\,\vec{A}_\mu\times\vec{\r}_\nu
  +\frac{e\cos\theta_C}{2\sin\theta_W}\,
  \vec{W}_\mu\times(\vec{\r}_\nu-\vec{a}_{1\,\nu})\, ,
\nonumber\\
D_\mu\vec{a}_{1\,\nu}&=
  \partial_\mu\,\vec{a}_{1\,\nu}+e\,\vec{A}_\mu\times\vec{a}_{1\,\nu}
  +\frac{e\cos\theta_C}{2\sin\theta_W}\,
  \vec{W}_\mu\times(\vec{a}_{1\,\nu}-\vec{\r}_\nu)\, .
\end{align}

One should mention that the replacements described above must be made
before the redefinition of the $\s$ field, \Eq{shiftsigma}. When we
now perform the shift of the $\sigma$ field, $W-\p$ and $W-\a$
couplings emerge. As a consequence there are two contributions to the
pion decay, a direct one and a contribution with an intermediate
longitudinal $a_1$:
\begin{equation}
f_\p = \sqrt{\zp{1}}
  \Big(\s_0-\frac{g^2\s_0^3}{\ma^2-\xi\,\mpii{1}^2}\Big)
  = \frac{\s_0\,(\ma^2-\xi\mpi^2-g^2\s_0^2)}
  {\sqrt{(\ma^2-\xi\mpi^2)^2-g^2 \s_0^2 \ma^2}}
\, .
\label{fpitree}
\end{equation}

Obviously we did not assume anything about vector meson dominance when
we constructed the coupling of the $A$ and $W$ fields to the mesons,
and in principle the vector mesons were not treated differently from
the scalar mesons. Hence, the $A$ and $W$ fields simply couple to the
Noether currents. In addition there are seagull terms involving two
$A$ or $W$ fields. At this stage there is no direct $\gamma-\r$ or
$W-\r$ coupling in our model. However, as we will see in
\Sect{Sectv1}, the $\gamma-\r$ and $W-\r$ coupling appears naturally
at the one-loop level, where we will need a counterterm of the form
\begin{equation}
\calL_{\gamma Y}+\calL_{W Y} = -\frac{f}{4}\,\tr
  (\partial_\mu Y_\nu-\partial_\nu Y_\mu)\,\partial^\mu\,\Big(e A^\nu T_3
  +\frac{e\cos\theta_C}{\sin\theta_W}\,(W_1^\nu T_1^L+W_2^\nu T_2^L)\Big)\, .
\label{LgammaYdirect}
\end{equation}
The $\gamma-\r$ part of this term has exactly the form of the
$\gamma-\r$ coupling of Kroll, Lee and Zumino \cite{KrollLeeZumino},
\begin{equation}
\calL_{\gamma\r} = -\frac{e f}{2}\,
  (\partial_\mu \r_{3\,\nu}-\partial_\nu\r_{3\,\mu})
  (\partial^\mu A^\nu-\partial^\nu A^\mu)\, ,
\end{equation}
the only difference being that in \Ref{KrollLeeZumino} the coupling
constant $f$ was fixed to $1/g$, whereas it is a free parameter in
our model. 
%
\section{Meson propagators in one-loop approximation}
%
\label{Sectoneloop}
Our aim is the chirally symmetric description of the vector isovector
and axial-vector isovector correlation functions. This implies that we
need a chirally symmetric description of $\r$ and $\a$ mesons. By
construction, our Lagrangian is chirally symmetric, but most
approximations beyond tree-level destroy the symmetry. For example, in
the linear $\s$ model without vector mesons the mean field
(Hartree-Fock) approximation violates the Goldstone theorem, which can
be recovered only if one includes also the RPA corrections
\cite{Aouissat}. Then, of course, the self-consistency is lost,
\ie the final RPA mesons are different from the mean-field mesons
propagating in the loops. In fact, a more general investigation of
this problem shows that there seems to be no self-consistent
approximation scheme which respects the symmetry at the correlator
level \cite{vanHees}.

A simple symmetry conserving approximation scheme is given by a loop 
expansion, which formally corresponds to an expansion in powers of $\hbar$. 
Here we will calculate the meson self energies to
one-loop order. This approximation has the following properties:
\begin{enumerate}
\item The pion remains massless in the chiral limit (Goldstone theorem).
\item All differences of chiral partners ($\p-\s$ and $\r-\a$)
  disappear when the chiral symmetry is restored ($\s_0=0$).
\item The dominant meson decay modes, such as $\r\rightarrow \p\p$ or
  $\a\rightarrow\p\r$, are contained in the scheme. Of course, due to the
  lack of self-consistency the final decay of the $\r$ meson in
  $\a\rightarrow\p\r\rightarrow \p\p\p$ is not included.
\end{enumerate}

Of course, in a strongly interacting theory the loop expansion
does not converge. Therefore the model is not defined by its
Lagrangian alone, but only by the Lagrangian together with the
approximation scheme used. Hence the propagators and vertices obtained
from the Lagrangian must not be interpreted as elementary, but rather
as parametrizations of the full propagators and vertices, containing
all non-perturbative effects which are not explicitly taken into
account in the approximation scheme.
%
\subsection{Subtraction of divergences}
%
\label{Sectdimreg}
If one calculates diagrams with loops, one always finds divergent
expressions which have to be regularized in some way. The above
mentioned fact, that the vertices and propagators obtained from our
Lagrangian are not elementary, justifies the introduction of form
factors, which are frequently used in the literature in order to cut
off high momenta \cite{ChanfraySchuck,ChanfrayRappWambach,Isgur},
rendering the loop integrals finite. These form factors are
independent parameters and cannot be computed from the
Lagrangian. Unfortunately we cannot simply attach form factors to the
vertices obtained from our Lagrangian, since this would destroy the
symmetry properties of our model.

Other regularization prescriptions, which can be found in the
literature, are the Pauli-Villars scheme with finite regulator mass
\cite{Herrmann} or subtracted dispersion relations. Using these
prescriptions one can sometimes avoid the problems arising from form
factors, although they give very similar results. As an example
consider the self-energy $\Sigma_{\r\rightarrow\p\p}$, describing the
decay of the $\r$ meson into two pions, with a three dimensional form
factor $v(\vec{k})$ attached to each $\rho\pi\pi$ vertex, which cuts
off pion momenta $\vec{k}$ (measured in the $\r$-meson rest frame) at
some value $|\vec{k}|_\mathrm{max}$\,. For small $\r$-meson energies,
$q^2\lesssim 4(\mpi^2+\vec{k}_\mathrm{max}^2)$, the imaginary part of
$\Sigma_{\r\rightarrow\p\p}$ is not modified by the form factor. For
the real part the situation is more complicated. Since it is related
to the imaginary part via a dispersion integral, it receives already
for low $q^2$ contributions from the imaginary part at high
$q^2$. Therefore it depends strongly on
$|\vec{k}|_\mathrm{max}$. However, using subtracted dispersion
relations, one can easily convince oneself that for low energies the
influence of the form factor on the real part can approximately be
absorbed in the subtraction constants, while the remaining dispersion
integral is almost independent of the form factor.

In this sense, subtracted dispersion relations can be interpreted as an
approximate way of attaching form factors to the vertices. 
As the subtraction constants depend on the high-energy behavior of the
form factors, they cannot be calculated from the Lagrangian, but they are
free parameters. However, there are some contraints which follow from
chiral symmetry. In order to preserve the symmetry, we must make equal
subtractions from the self energies of chiral partners. Obviously this
is mathematically ill-defined, since the subtractions are
infinite. Therefore we need a more systematic method to ensure the
consistency of the subtractions.

The subtractions can be regarded as (infinite)
counterterms added to the original Lagrangian. Thus the subtractions
are consistent with chiral symmetry if the counterterms are chirally
invariant. The values of the counterterms are determined as follows
\cite{RamMohan}. First we calculate the self-energy diagrams with
dimensional regularization, \ie we calculate the loop integrals in
$d$ space-time dimensions,
\begin{equation}
\intd{4}{k}\cdots\longrightarrow
\intd{d}{k}\Big(\frac{\Lambda^2}{4\p}\Big)^{\frac{4-d}{2}}
  \,\cdots\, .
\label{dimregul}
\end{equation}
If we expand the result in the vicinity of $d=4$, the divergences
appear as poles $\propto 1/(4-d)$. The residues of these poles are
mathematically well-defined, and the self energy can be written in the
form
\begin{equation}
\Sigma = \overline{\Sigma} +
  \Sigma_\infty\,\Big(\frac{1}{4-d}-\frac{\gamma}{2}\Big)\, .
\end{equation}
In this way we have split the self energy into a finite part
$\overline{\Sigma}$, and an infinite part. ($\gamma = 0.577\ldots=$
Euler's constant is finite, but it is convenient to keep the infinite
$1/(4-d)$ term and the $\gamma/2$ term together.) The counterterms are
also split into finite and infinite parts, \eg $\delta\mu^2 =
\ovd{\mu^2}+\delta \mu^2_\infty(1/(4-d)-\gamma/2)$. The infinite parts
of the counterterms must be adjusted such that $\Sigma_\infty$
vanishes, while the finite parts are treated as free parameters. The
crucial point is that each counterterm may be adjusted only once and
not separately for different self energies.

The factor $(\Lambda^2/(4\p))^{(4-d)/2}$ in \Eq{dimregul} ensures that
the dimension of the result does not depend on $d$. The finite parts
of the integrals, of course, depend on the choice of
$\Lambda$. However, the physical results are independent of $\Lambda$,
if the finite parts of the counterterms are appropriately readjusted
after a change of $\Lambda$. For example, if $\Lambda$ is changed to
$\Lambda^\prime$, $\ovd{\mu^2}$ must be changed to $\ovd{\mu^2} +
\delta \mu^2_\infty \ln(\Lambda^\prime/\Lambda)$.
%
\subsection{The tadpole correction}
\begin{figure}
\begin{center}
\epsfig{file=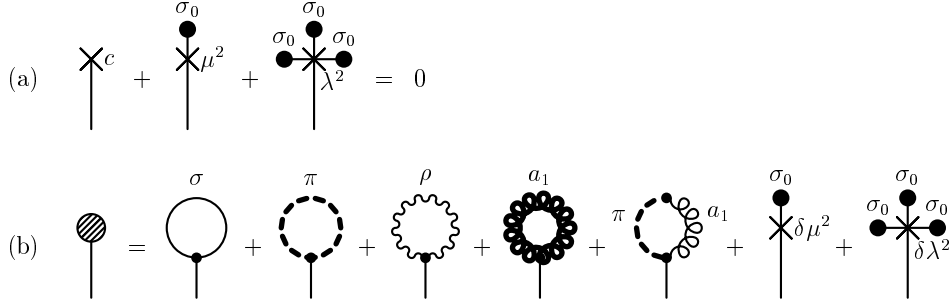,width=12.6cm,bbllx=84,bblly=618,bburx=528,bbury=760}
\end{center}
\vspace{-0.5cm}
\caption{\small (a) Diagrammatic illustration of \Eq{minimumsigma0},
  which shows that there are no tadpole graphs at tree level. (b)
  Tadpole graphs $T$ at one-loop order. The meaning of thick and thin
  lines is the same as in \Fig{Figpia1mix}.}
\label{Figtadpole}
\end{figure}

%
At tree level, no $\s$ line can disappear into the vacuum, because all
terms linear in $\s$ are eliminated by the shift of the $\s$ field,
\Eq{shiftsigma}, as a consequence of \Eq{minimumsigma0}, which
can be illustrated diagrammatically as shown in \Fig{Figtadpole}(a).
This is no longer true at the one-loop level, where one has the
additional tadpole graphs shown in \Fig{Figtadpole}(b). The last two
contributions are generated by the chirally symmetric counterterms
\begin{equation}
\delta\calL_{\mu^2}+\delta\calL_{\lambda^2}=
-\frac{\delta\mu^2}{2}\Phi\cdot\Phi-\frac{\delta\lambda^2}{4}(\Phi\cdot\Phi)^2
\, .
\label{dLmudLlambda}
\end{equation}
Calculating the diagrams displayed in \Fig{Figtadpole}(b) and
splitting the result into finite and infinite parts,
$T=\overline{T}+T_\infty (1/(4-d)-\gamma/2)$, we find for the infinite
part
\begin{align}
T_\infty=\,&\frac{3i\s_0}{8\p^2}\Big(\frac{(h_1+2 h_2-g^2)m_0^2}{\xi^2}
  -\frac{g^2\mu^2}{\xi}+(3 h_1+6 h_2)m_0^2+2\lambda^2\mu^2\Big)
\nonumber\\
& +\frac{3i\s_0^3}{8\p^2}\Big(\frac{(h_1+h_2-g^2)^2+h_2^2}{\xi^2}
  -\frac{2 g^2(h_1+h_2)}{\xi}+3(h_1+h_2)^2+3h_2^2+4\lambda^2\Big)
\nonumber\\
& -i\s_0\delta\mu^2_\infty-i\s_0^3\delta\lambda^2_\infty\, .
\label{tadpolei}
\end{align}

From this one can immediately see that $T$ can be made finite (\ie
$T_\infty=0$) by a proper choice of $\delta\mu^2_\infty$ and
$\delta\lambda^2_\infty$\,. Furthermore, this choice does not depend
on $c$ or $\s_0$, \ie on the explicit or spontaneous symmetry
breaking. In the following sections we will see that this is a general
property of all counterterms, as shown by Lee \cite{Lee} for the
linear $\s$ model without vector mesons and by Ram Mohan
\cite{RamMohan} for the linear $\s$ model with nucleons at finite
temperature and density. In this context we should remark that, in the
limit $\xi\rightarrow 0$, it is impossible to find suitable
counterterms independent of $\s_0$\,. For example, the third
contribution in \Fig{Figtadpole}(b) can be written as
\begin{align}
-3 h_2\,\s_0\,g_{\mu\nu}\intd{4}{k} G_\r^{\mu\nu}(k)
&=3 h_2\,\s_0\intd{4}{k}
  \Big(\frac{3}{k^2-\mr^2}+\frac{1}{\xi k^2-\mr^2}\Big)
\nonumber\\
&=3\sum_{n=1}^{\infty} h_2^n\,\s_0^{2n-1}\intd{4}{k}
  \Big(\frac{3}{(k^2-m_0^2)^n}+\frac{1}{(\xi k^2-m_0^2)^n}\Big)\, .
\end{align}
For $\xi>0$ the $n=1$ and $n=2$ terms are quadratically and
logarithmically divergent, respectively, while all terms with $n>2$
are finite. This is the reason why two counterterms independent of
$\s_0$ are sufficient to cancel the divergences. In contrast, for
$\xi=0$, all terms are quartically divergent, \ie one needs either an
infinite number of counterterms independent of $\s_0$\,, or
counterterms which depend on $\s_0$\,.
%
\subsection{The \boldmath{$\r$}-meson self energy}
\begin{figure}
\begin{center}
\epsfig{file=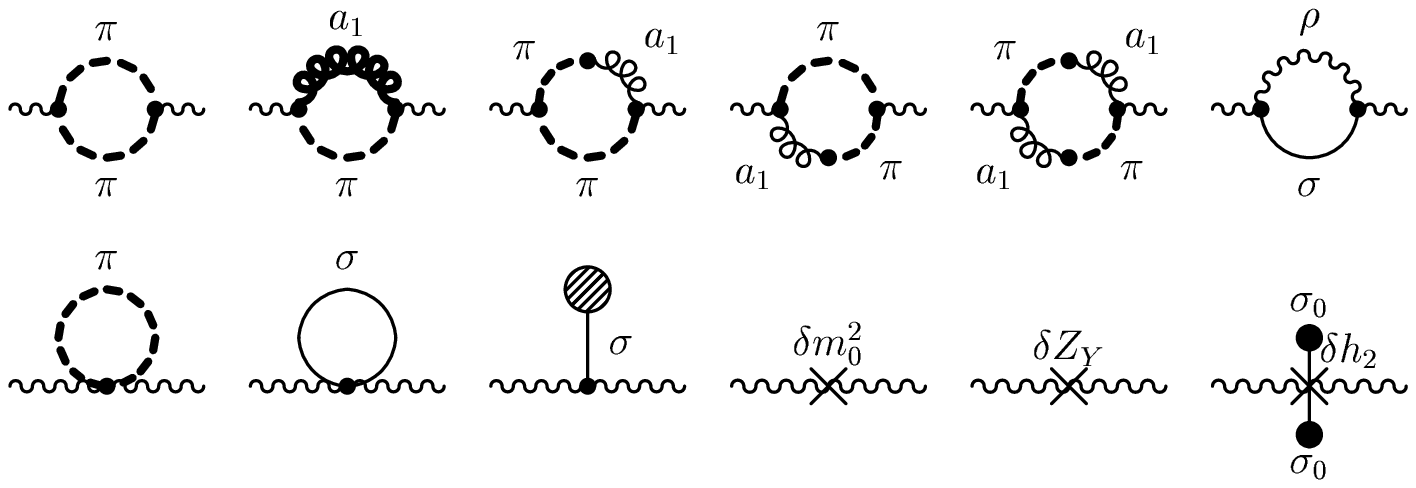,width=11.6cm,bbllx=69,bblly=649,bburx=476,bbury=787}
\end{center}
\vspace{-0.5cm}
\caption{\small One-loop contributions to the $\r$-meson self energy,
  $-i\Sigma_\r$\,.}
\label{FigSigrho}
\end{figure}

%
In \Fig{FigSigrho} all one-loop diagrams for the $\r$-meson
self energy are displayed.
The last three diagrams correspond to the contributions of the three
counterterms which are needed to cancel the divergences:
\begin{align}
\delta\calL_{m_0^2}+\delta\calL_{Z_Y}+\delta\calL_{h_2}=\,
& \frac{\delta m_0^2}{4}\,\tr Y_\mu Y^\mu
  +\frac{\delta Z_Y}{8}\,\tr(\partial_\mu Y_\nu-\partial_\nu Y_\mu)
  (\partial^\mu Y^\nu-\partial^\nu Y^\mu)
\nonumber\\
& +\frac{\delta h_2}{4}\,\Phi\cdot\Phi\,\tr Y_\mu Y^\mu 
\label{dLm02dLZYdLh2}
\end{align}
Again, the self energy is split into finite and infinite parts,
$\Sigma_\r^{\mu\nu} = \overline{\Sigma}_\r^{\mu\nu} +
\Sigma_{\r\infty}^{\mu\nu} (1/(4-d)-\gamma/2)$, and the infinite parts
of the counterterms are chosen such that $\Sigma_{\r\infty}^{\mu\nu}$
vanishes%
\footnote{\samepage{For non-vanishing vector-meson self interaction, \ie
  $g^{\prime}\neq 0$ or $g^{\prime\prime}\neq 0$, there are more diagrams
  than shown in \Fig{FigSigrho}, and one additional counterterm is
  needed:
  \begin{equation}
    \delta\calL_{\xi} =
      -\frac{\delta\xi}{4}\,\tr (\partial_\mu Y^\mu)^2\, .
  \end{equation}}}
.

Since in our model the $\r$ meson is not a gauge boson, the $\r$-meson
self energy is not transverse. Instead, it can be decomposed into
transverse and longitudinal components,
\begin{equation}
\Sigma_\r^{\mu\nu}(q)=
  \Sigma_\r^t(q^2)\Big(\frac{q^\mu q^\nu}{q^2}-g^{\mu\nu}\Big)
  +\Sigma_\r^l(q^2)\frac{q^\mu q^\nu}{q^2}\, .
\label{Sigmatl}
\end{equation}
As we will see, the longitudinal part does not contribute to the
vector current correlator as a consequence of current conservation.
%
\subsection{The pion and \boldmath{$\a$}-meson self energies}
%
Because of the $\p-\a$ mixing, the situation for the pion is a bit
more complicated than for the $\r$ meson. It is necessary to
distinguish between the one-particle irreducible (1PI) self energy
$\Sigma_{\p\p}$\,, which cannot be cut into two parts by cutting any
single line, and the proper self energy $\Sigma_\p$\,, which cannot be
cut into two parts by cutting a single pion line, but which can
possibly be cut into two parts by cutting a single $\a$-meson line
(see \eg the last diagram in \Fig{Figpia1mix}(a)).

For the $\a$ meson the situation is similar. Here we have to
distinguish between the 1PI self energy $\Sigma_{\a\a}$\,, which
cannot be cut into two parts by cutting any single line, and the
proper self energy $\Sigma_{\a}$\,, which cannot be cut into two parts
by cutting a single $\a$-meson line, but which can possibly be cut
into two parts by cutting a single pion line.

\begin{figure}
\begin{center}
\epsfig{file=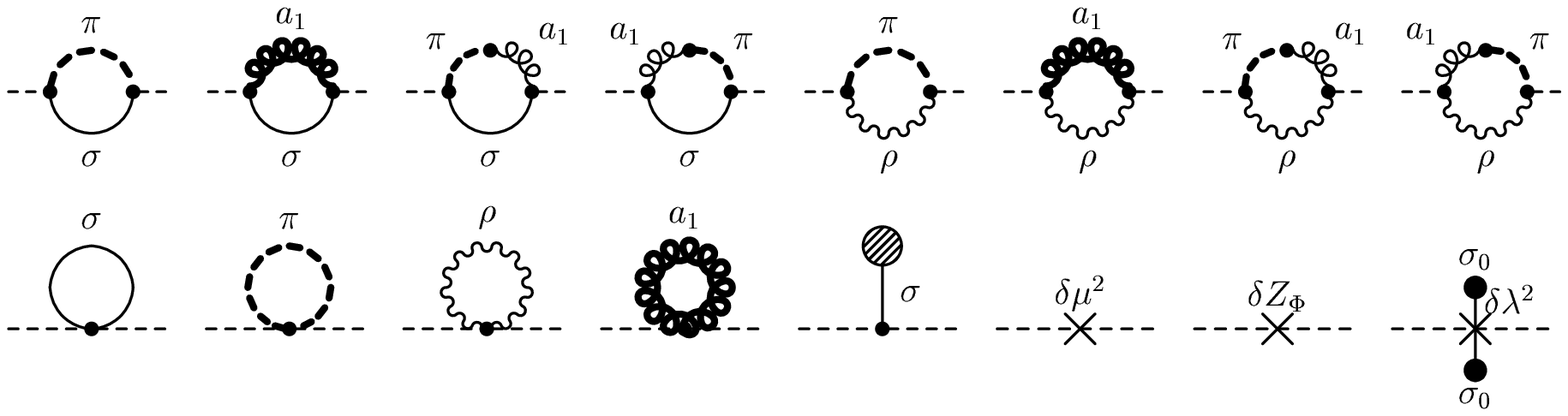,width=14.8cm,bbllx=33,bblly=649,bburx=555,bbury=786}
\end{center}
\vspace{-0.5cm}
\caption{\small One-loop contributions to the 1PI pion self energy,
  $-i\Sigma_{\p\p}$\,.}
\label{FigSigpipi}
\end{figure}

In \Fig{FigSigpipi} we display the one-loop contributions to the 1PI pion
self energy $\Sigma_{\p\p}$\,. The last three diagrams are generated
by the counterterms given in \Eq{dLmudLlambda} and the additional
wave-function renormalization counterterm,
\begin{equation}
\delta\calL_{Z_\Phi}=
-\frac{\delta Z_\Phi}{2}\partial_\mu\Phi\cdot\partial^\mu\Phi\, .
\label{dLZPhi}
\end{equation}
One can explicitly show that this self energy fulfills the Goldstone
theorem, \ie
\begin{equation}
\Sigma_{\p\p}(0)=0\quad\mathrm{for}\quad c=0\, .
\label{Goldstone1loop}
\end{equation}
In particular, it is interesting to see that in the chiral limit the
contributions from the $\delta\mu^2$ and $\delta\lambda^2$
counterterms ($14^\mathrm{th}$ and $16^\mathrm{th}$ diagram) exactly
cancel the counterterm contributions contained in the tadpole
correction ($13^\mathrm{th}$ diagram).

\begin{figure}
\begin{center}
\epsfig{file=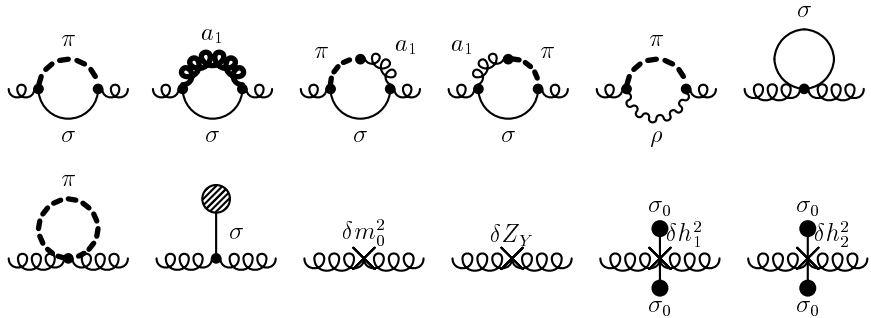,width=11.5cm,bbllx=48,bblly=634,bburx=453,bbury=783}
\end{center}
\vspace{-0.5cm}
\caption{\small One-loop contributions to the 1PI $\a$-meson self
  energy, $-i\Sigma_{\a\a}$\,.}
\label{FigSiga1}
\end{figure}

The one-loop diagrams for the 1PI $\a$-meson self energy
$\Sigma_{\a\a}^{\mu\nu}$ are displayed in \Fig{FigSiga1}.
The diagrams 9, 10 and 12 are generated by the counterterms given in
\Eq{dLm02dLZYdLh2}. In addition we need one more counterterm, namely
\begin{equation}
\delta\calL_{h_1} = -\frac{\delta h_1}{2}\,Y_\mu \Phi\cdot Y^\mu \Phi\, ,
\end{equation}
which gives rise to the $11^\mathrm{th}$ diagram. In analogy to the
$\r$-meson self energy (cf. \Eq{Sigmatl}), the $\a$-meson self energy
can be decomposed into transverse and longitudinal components,
$\Sigma_{\a\a}^t$ and $\Sigma_{\a\a}^l$. The transverse part is not
modified by $\p-\a$ mixing, and therefore the proper transverse
$\a$-meson self energy is equal to the 1PI transverse self energy,
\begin{equation}
\Sigma_{\a}^t(q^2)=\Sigma_{\a\a}^t(q^2)\, .
\end{equation}

\begin{figure}
\begin{center}
\epsfig{file=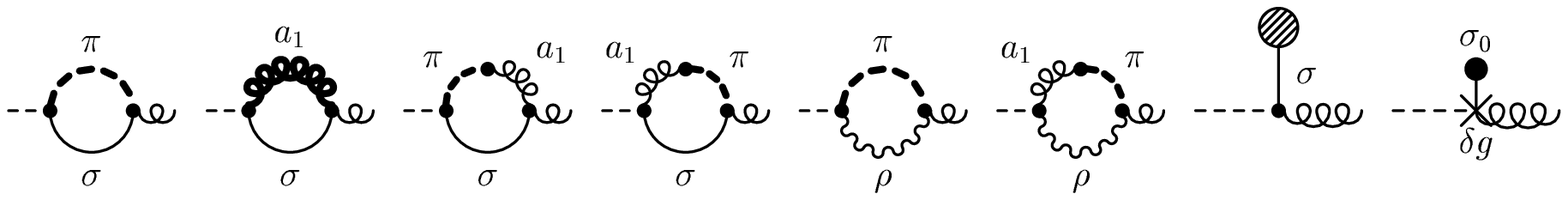,width=15cm,bbllx=32,bblly=713,bburx=560,bbury=779}
\end{center}
\vspace{-0.5cm}
\caption{\small One-loop diagrams for the 1PI mixed $\p-\a$
  self energy, $-i\Sigma_{\p\a}^\mu$.}
\label{FigSigpia1}
\end{figure}

To obtain the proper pion self energy, we have to consider also the
1PI mixed self energy $\Sigma_{\p\a}^\mu$, which is a correction to
the $\p-\a$ vertex given by the Lagrangian (\ref{Lpia1mix}). The
corresponding one-loop diagrams are shown in \Fig{FigSigpia1}.
The last diagram comes from the counterterm
\begin{equation}
\delta\calL_g = -i \delta g\, Y_\mu \Phi\cdot\partial^\mu\Phi\, .
\label{dLg}
\end{equation}
With this mixed self energy, the corrected $\p-\a$ mixing vertex is
given by
\begin{equation}
\big(\Gamma_{\p\a}(q)\big)_{ab}^{\mu}=
  -\delta_{ab}\big(g\s_0\,q^{\mu}+i\Sigma_{\p\a}^\mu(q)\big) =:
  -q^\mu \delta_{ab}\, F_{\p\a}(q^2)\, ,
\end{equation} 
where $q$ is the incoming momentum of the pion. The proper pion self
energy contains the tree-level self energy shown in
\Fig{Figpia1mix}(a) and the loop corrections, \ie
\begin{equation}
\Sigma_\p(q^2) = \Sigma_{\p\p}(q^2)
  +\frac{q^2 \big(F_{\p\a}(q^2)\big)^2}
    {\ma^2-\xi q^2-\Sigma_{\a\a}^l(q^2)}\, .
\end{equation}
The corrected pion propagator is given by $G_\p^{(1)} =
1/(q^2-\mpii{0}^2-\Sigma_\p)$. It is obvious that $\Sigma_\p$ fulfills
the Goldstone theorem if $\Sigma_{\p\p}$ does, because the
contributions due to $\p-\a$ mixing are proportional to $q^2$.
%
\subsection{The \boldmath{$\s$}-meson self energy}
\begin{figure}
\begin{center}
\epsfig{file=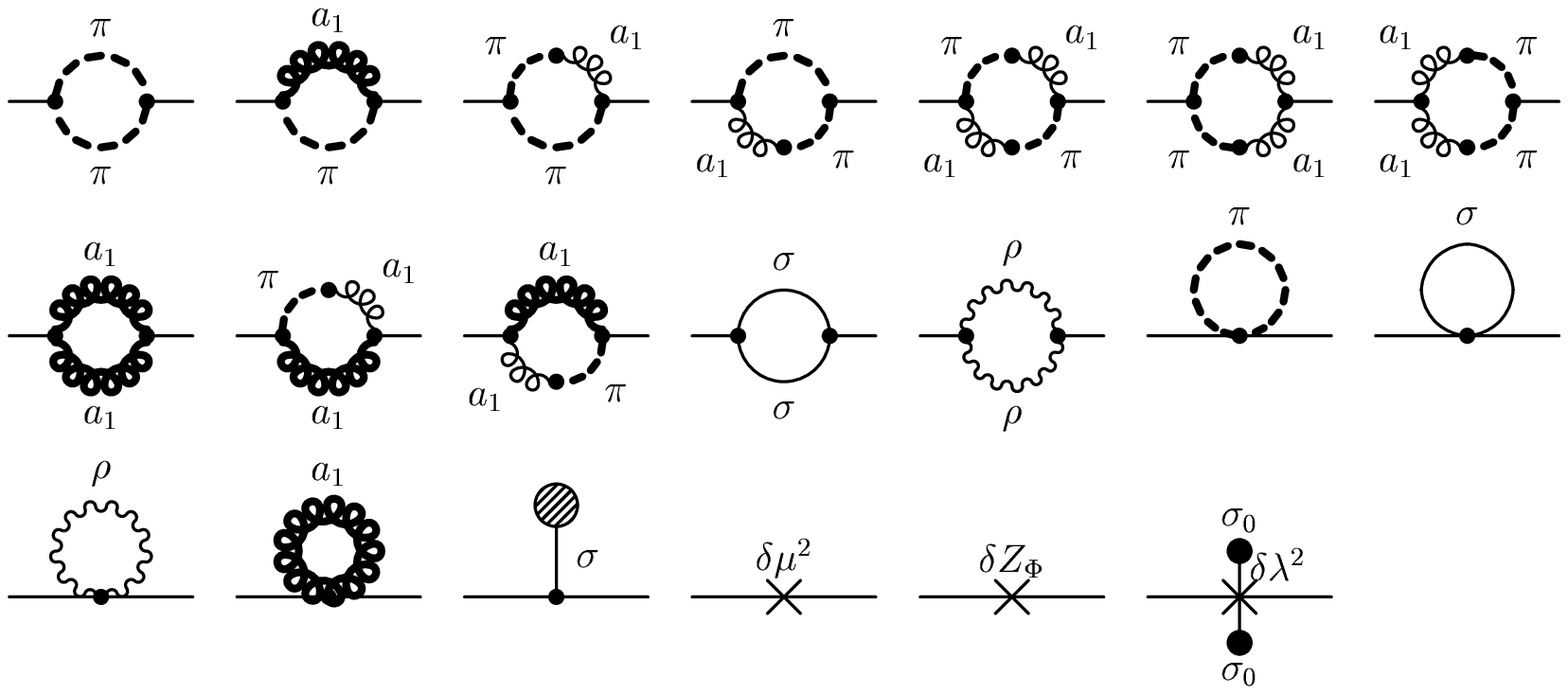,width=13.5cm,bbllx=54,bblly=577,bburx=530,bbury=786}
\end{center}
\vspace{-0.5cm}
\caption{\small One-loop contributions to the $\s$-meson self energy,
  $-i\Sigma_\s$\,.}
\label{FigSigsig}
\end{figure}

%
The self energy of the $\s$-meson, $\Sigma_\s$, is given by the
diagrams shown in \Fig{FigSigsig}.
To make it finite, the counterterms $\delta\mu^2$, $\delta\lambda^2$
and $\delta Z_\Phi$, introduced already in the previous sections, are
sufficient.
%
\section{Vector and axial-vector correlation functions}
%
\label{Sectcorrelators}
Let $\vec{j}_\mu$ and $\vec{j}^5_\mu$ be the hadronic vector isovector
and axial-vector isovector currents, respectively. In terms of quark
fields $\psi=(u,d)$ these currents read $\vec{j}_\mu =
\overline{\psi}\gamma_\mu\vec{\tau}\psi/2$ and $\vec{j}^5_\mu =
\overline{\psi}\gamma_\mu\gamma_5\vec{\tau}\psi/2$. The
current-current correlation functions are defined by
\begin{align}
\Pi_{V\,ab}^{\mu\nu}(q)&=i\hspace{-1mm}\int\hspace{-2mm}d^4x\,e^{iq\cdot x}
\langle 0\vert T\big(j_a^\mu(x) j_b^\nu(0)\big)\vert 0\rangle
\, ,\label{DefPiV}\\
\Pi_{A\,ab}^{\mu\nu}(q)&=i\hspace{-1mm}\int\hspace{-2mm}d^4x\,e^{i q\cdot x}
\langle 0\vert T\big(j_a^{5\,\mu}(x) j_b^{5\,\nu}(0)\big)\vert 0\rangle
\, .\label{DefPiA}
\end{align}
Because of isospin symmetry we have
$\Pi_{V/A\,ab}^{\mu\nu}=\Pi_{V/A}^{\mu\nu}\,\delta_{ab}$\,. The
correlation functions can be split into longitudinal and transverse
components as described for the $\r$- and $\a$-meson self energies:
\begin{equation}
\Pi_{V/A}^{\mu\nu}(q)=\Pi_{V/A}^t(q^2)
  \Big(\frac{q^\mu q^\nu}{q^2}-g^{\mu\nu}\Big)
  +\Pi_{V/A}^l(q^2)\frac{q^\mu q^\nu}{q^2}\, .
\end{equation}
Since the vector current is conserved, its longitudinal correlator
vanishes, \ie $\Pi_V^l = 0$\,. As usual, inserting a complete set of
energy and momentum eigenstates into \Eqs{DefPiV} and (\ref{DefPiA}),
one can derive spectral representations for the correlation
functions. Following the definitions of \Ref{ALEPH}, the corresponding
spectral functions $v_1$, $a_1$ and $a_0$ are given by
\begin{equation}
v_1(q^2)=\frac{4\p}{q^2} \Im\Pi_V^t(q^2)\, ,\quad
a_1(q^2)=\frac{4\p}{q^2} \Im\Pi_A^t(q^2)\, ,\quad
a_0(q^2)=\frac{4\p}{q^2} \Im\Pi_A^l(q^2)\, .
\label{Defv1a1a0}
\end{equation}
To order $e^2$, the current-current correlation functions are directly
related to the photon and $W$-boson self energies:
\begin{align}
\Sigma_\gamma^{\mu\nu}&=-e^2\Pi_V^{\mu\nu}
  +\mbox{isoscalar contributions}\, ,
\label{SiggamPiV}
\\
\Sigma_W^{\mu\nu}&=
  -\Big(\frac{e \cos\theta_C}{2\sin\theta_W}\Big)^2
  (\Pi_V^{\mu\nu}+\Pi_A^{\mu\nu})\, .
\label{SigWPiA}
\end{align}
Experimentally, $\Pi_V$ can be measured in the reaction
$e^++e^-\rightarrow$ hadrons. Up to $q^2=m_\tau^2$\,, both $\Pi_V$ and
$\Pi_A$ can be measured in the reaction $\tau\rightarrow\nu_\tau+$
hadrons.
%
\subsection{Photon self energy and vector correlator}
%
\label{Sectv1}
In this section we will calculate the spectral function $v_1$ of the
vector correlator, which is proportional to the imaginary part of the
(isovector) photon self energy $\Sigma_\gamma$. Similar to the pion
and $\a$-meson self energies, the proper photon self energy
$\Sigma_\gamma$ has a one-particle irreducible (1PI) contribution,
$\Sigma_{\gamma\gamma}$, and contributions which can be cut into two
parts by cutting a single line, which in this case must be a
$\r$-meson line.

\begin{figure}
\begin{center}
\epsfig{file=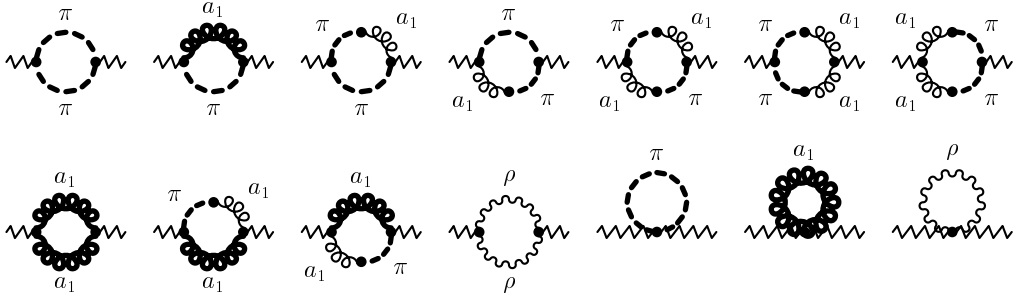,width=13.5cm,bbllx=55,bblly=646,bburx=531,bbury=785}
\end{center}
\vspace{-0.5cm}
\caption{\small One-loop diagrams for the 1PI photon self energy,
  $-i\Sigma_{\gamma\gamma}^{\mu\nu}$.}
\label{FigSiggamgam}
\end{figure}

In \Fig{FigSiggamgam} we display the one-loop diagrams for the 1PI
photon self energy, $\Sigma_{\gamma\gamma}^{\mu\nu}$\,.
Here we do not introduce counterterms to cancel the divergences,
because we are only interested in the imaginary part, which is
finite. The last three diagrams are purely real and do not contribute
to the spectral function $v_1$, but taking them into account one can
explicitly show that within the dimensional regularization scheme the
self energy is gauge invariant, \ie
\begin{equation}
q_\mu \Sigma_{\gamma\gamma}^{\mu\nu}(q) = 0
\qquad\mathrm{and}\qquad
\Sigma_{\gamma\gamma}^{\mu\nu}(0) = 0\, .
\label{gaugeinvSgg}
\end{equation}

\begin{figure}
\begin{center}
\epsfig{file=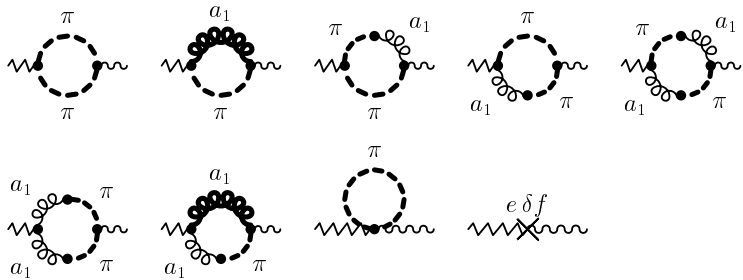,width=9.9cm,bbllx=67,bblly=654,bburx=417,bbury=785}
\end{center}
\vspace{-0.5cm}
\caption{\small One-loop diagrams for the mixed $\gamma-\r$ self energy,
  $-i\Sigma_{\gamma\r}^{\mu\nu}$.}
\label{FigSiggamrho}
\end{figure}

For the remaining self energy contributions we must compute the
$\gamma-\r$ transition vertex. The corresponding mixed self energy,
$\Sigma_{\gamma\r}^{\mu\nu}(q)$, is given by the diagrams shown in
\Fig{FigSiggamrho}.
Again, the conditions for gauge invariance,
\begin{equation}
q_\mu \Sigma_{\r\gamma}^{\mu\nu}(q) = 0
\qquad\mathrm{and}\qquad
\Sigma_{\r\gamma}^{\mu\nu}(0) = 0\, ,
\label{gaugeinvSgr}
\end{equation}
are fulfilled. This is the reason why only one counterterm is needed
to cancel the remaining divergence, namely a counterterm of the same
form as the direct gauge-boson $-$ vector-meson coupling,
\Eq{LgammaYdirect},
\begin{equation}
 \delta\calL_f = -\frac{\delta f}{4}\,\tr
  (\partial_\mu Y_\nu-\partial_\nu Y_\mu)\,\partial^\mu\,\Big(e A^\nu T_3
  +\frac{e\cos\theta_C}{\sin\theta_W}\,(W_1^\nu T_1^L+W_2^\nu T_2^L)\Big)\, .
\label{dLf}
\end{equation}
In terms of the mixed self energy, the one-loop corrected $\gamma-\r$
vertex function can be written as
\begin{equation}
\big(\Gamma_{\gamma\r}(q)\big)_a^{\mu\nu} 
  = i \delta_{3a} \big(e f (q^\mu q^\nu-q^2 g^{\mu\nu})
    -\Sigma_{\gamma\r}^{\mu\nu}(q)\big)
  =: i e \delta_{3a}(q^\mu q^\nu-q^2 g^{\mu\nu})\,F_{\gamma\r}(q^2)\, .
\end{equation}
Here $a$ is the isospin index of the $\r$ meson.

Now we can write down the total photon self energy. Because of
\Eqs{gaugeinvSgg} and (\ref{gaugeinvSgr}) its longitudinal component
vanishes. The transverse component reads
\begin{equation}
\Sigma_\gamma^t(q^2) = \Sigma_{\gamma\gamma}^t(q^2)
  +\frac{\big(e q^2 F_{\gamma\r}(q^2)\big)^2}
    {q^2-\mr^2-\Sigma_\r^t(q^2)}\, .
\end{equation}
Finally we obtain the vector spectral function $v_1(q^2)$ by taking
the imaginary part of $\Sigma_\gamma^t$, see \Eqs{Defv1a1a0} and
(\ref{SiggamPiV}).
%
\subsection{Transverse part of the axial-vector correlator}
%
Similar to the vector correlator, the axial-vector correlator has
one-particle irreducible contributions, and contributions which can be
cut into two parts by cutting a single line, in this case either an
$\a$-meson or a pion line. Of course, the single-pion states can
contribute only to the longitudinal component of the correlator, which
will be postponed to the next section.

\begin{figure}
\begin{center}
\epsfig{file=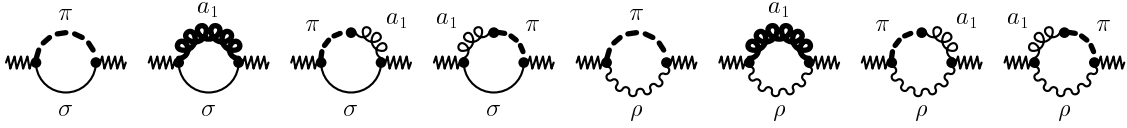,width=14.9cm,bbllx=33,bblly=726,bburx=555,bbury=786}
\end{center}
\vspace{-0.5cm}
\caption{\small One-loop diagrams for the 1PI axial-vector $W$-boson
  self energy, $-i\Sigma_{WW}^{A\,\mu\nu}$ (seagull and tadpole graphs
  omitted).}
\label{FigSigWW}
\end{figure}

Due to the parity violation of weak interactions, the $W$-boson self
energy has vector and axial-vector contributions, $\Sigma_W =
\Sigma_W^V + \Sigma_W^A$\,. In this section we are looking only at the
axial-vector contribution, $\Sigma_W^A$. Its one-particle irreducible
(1PI) part, $\Sigma_{WW}^A$, is given by the diagrams shown in
\Fig{FigSigWW} and some real seagull and tadpole contributions, which
we have omitted since we are only interested in the imaginary part.
As in the previous section, the imaginary part is finite. Since the
axial current is not conserved, $\Sigma_{WW}^{A\,\mu\nu}$ is not
transverse and can be split into transverse and longitudinal
components, $\Sigma_{WW}^{A\,t}$ and $\Sigma_{WW}^{A\,l}$.

\begin{figure}
\begin{center}
\epsfig{file=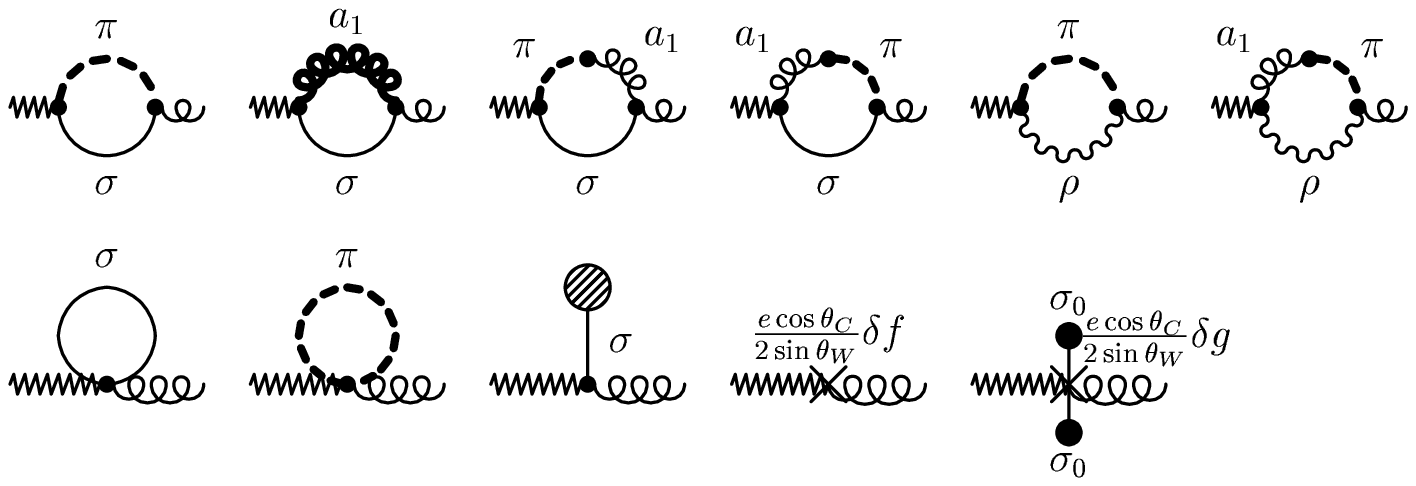,width=11.6cm,bbllx=69,bblly=648,bburx=476,bbury=786}
\end{center}
\vspace{-0.5cm}
\caption{\small One-loop diagrams for the mixed $W-\a$ self energy,
  $-i\Sigma_{Wa1}^{\mu\nu}$\,.}
\label{FigSigWa1}
\end{figure}

In \Fig{FigSigWa1} the mixed $W-\a$ self energy,
$\Sigma_{W\a}^{\mu\nu}$, is shown.
The missing current conservation does not only lead to a non-vanishing
longitudinal component, $\Sigma_{W\a}^l$, it also avoids the
cancellation of divergences due to the condition
$\Sigma_{W\a}^{\mu\nu}(0)=0$, which would be valid if the current was
conserved. Hence, the single $\delta f$ counterterm given in \Eq{dLf}
is not sufficient to render $\Sigma_{W\a}^{\mu\nu}$ finite. However,
the minimal substitution, \Eq{DPhi}, must also be applied to the
$\delta g$ counterterm introduced in \Eq{dLg}. This finally leads to
the last diagram in \Fig{FigSigWa1}, and indeed, with this counterterm
the remaining divergence in $\Sigma_{W\a}^{\mu\nu}$ is cancelled. The
total 1-loop corrected $W-\a$ vertex can now be written as
\begin{align}
\big(\Gamma_{W\a}(q)\big)_{ab}^{\mu\nu}
  &= -i\delta_{ab}\Big(\frac{e\cos\theta_C}{2\sin\theta_W}\big(
    f (q^\mu q^\nu-q^2 g^{\mu\nu}\big)+g\s_0^2 g^{\mu\nu}\big)
    +\Sigma_{W\a}^{\mu\nu}\Big)\nonumber\\
  &=: -i\delta_{ab}\frac{e\cos\theta_C}{2\sin\theta_W}\Big(
    \Big(\frac{q^\mu q^\nu}{q^2}-g^{\mu\nu}\Big) F_{W\a}^t(q^2)
    +\frac{q^\mu q^\nu}{q^2} F_{W\a}^l(q^2)\Big)\, .
\end{align}
Here $a$ is the index of the $W$ field (1 or 2) and $b$ is the isospin
index of the $\a$ meson.

The spectral function of the transverse axial correlator, $a_1(q^2)$,
is now obtained from
\begin{equation}
\Sigma_W^{A\,t}(q^2) = \Sigma_{WW}^{A\,t}(q^2)
  +\Big(\frac{e\cos\theta_C}{2\sin\theta_W}\Big)^2
  \frac{\big(F_{W\a}^t(q^2)\big)^2}{q^2-\ma^2-\Sigma_{\a}^t}\, ,
\end{equation}
together with \Eqs{Defv1a1a0} and (\ref{SigWPiA}).
%
\subsection{Longitudinal part of the axial-vector correlator}
\begin{figure}
\begin{center}
\epsfig{file=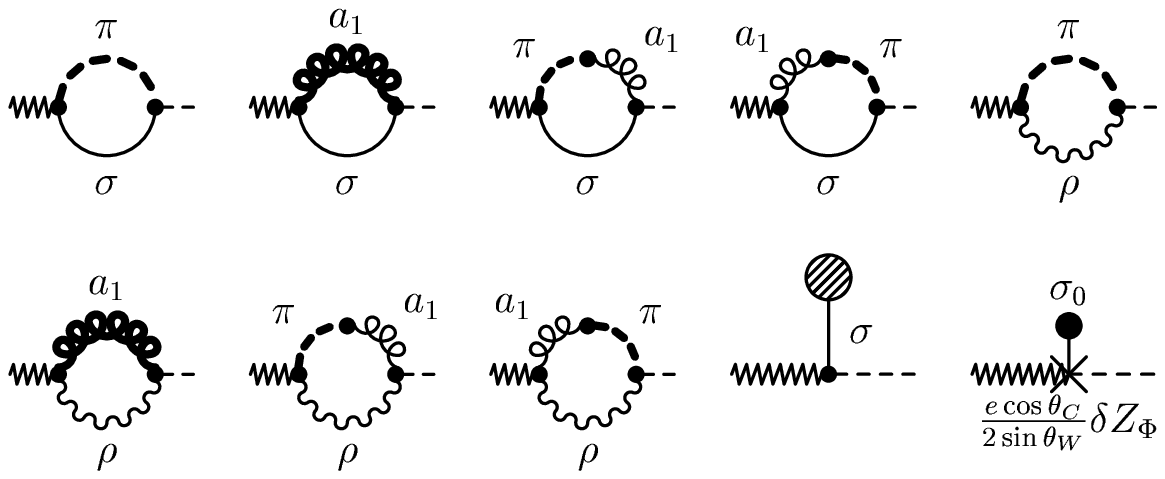,width=9.5cm,bbllx=68,bblly=649,bburx=404,bbury=786}
\end{center}
\vspace{-0.5cm}
\caption{\small One-loop diagrams for the mixed $W-\p$ self energy,
  $-i\Sigma_{W\p}^\mu$\,.}
\label{FigSigWpi}
\end{figure}

%
For the longitudinal part of the axial-vector correlator, we must also
consider contributions with single-pion intermediate states. To that
end we need the $W-\p$ transition vertex, which again consists of a
one-particle irreducible (1PI) contribution and a contribution which
can be cut into two parts by cutting a single $\a$-meson line. The 1PI
diagrams for the mixed $W-\p$ self energy, $\Sigma_{W\p}^\mu$, are
displayed in \Fig{FigSigWpi}.
The last diagram is generated by the counterterm one obtains by
performing the minimal substitution given by \Eq{DPhi} in the
wave-function renormalization counterterm, \Eq{dLZPhi}. The 1PI
one-loop corrected $W-\p$ vertex can now be written as
\begin{equation}
\big(\Gamma_{W\p}(q)\big)_{ab}^\mu =
  -\delta_{ab}\Big(\frac{e\cos\theta_C}{2\sin\theta_W}\s_0\,q^\mu
  +i\Sigma_{W\p}^\mu(q)\Big) =:
  -\frac{e\cos\theta_C}{2\sin\theta_W}q^\mu\delta_{ab}F_{W\p}(q^2)\, ,
\end{equation}
with $q$ being the incoming momentum of the $W$ boson.
Then the full $W-\p$ vertex reads
\begin{align}
\big(\Gamma_{W\p}^\mathrm{tot}(q)\big)_{ab}^\mu
&=-\frac{e\cos\theta_C}{2\sin\theta_W}q^\mu\delta_{ab}
  \Big(F_{W\p}(q^2)-\frac{F_{W\a}^l(q^2) F_{\p\a}(q^2)}
    {\ma^2-\xi q^2-\Sigma_{\a\a}^l(q^2)}\Big)
\nonumber\\
&=:-\frac{e\cos\theta_C}{2\sin\theta_W}q^\mu\delta_{ab}
  F_{W\p}^\mathrm{tot}(q^2)\, .
\end{align}
The longitudinal $W$-boson self energy, which determines the spectral
function $a_0(q^2)$ according to \Eqs{Defv1a1a0} and (\ref{SigWPiA}),
is given by
\begin{equation}
\Sigma_W^{A\,l}(q^2) = \Sigma_{WW}^{A\,l}(q^2)
  +\Big(\frac{e\cos\theta_C}{2\sin\theta_W}\Big)^2
    \Big(\frac{\big(F_{W\a}^l(q^2)\big)^2}
      {\ma^2-\xi q^2-\Sigma_{\a\a}^l(q^2)}
      +\frac{q^2\big(F_{W\p}^\mathrm{tot}(q^2)\big)^2}
        {q^2-\mpii{0}^2-\Sigma_\p(q^2)}\Big)\, .
\label{SigWl}
\end{equation}

The full $W-\p$ vertex can also be used to calculate the one-loop
corrected pion decay constant,
\begin{equation}
f_\p^{(1)} = F_{W\p}^\mathrm{tot}(\mpi^{(1)\,2})\;\sqrt{Z_\p^{(1)}}\, ,
\label{fpi1}
\end{equation}
with $\mpi^{(1)}$ and $Z_\p^{(1)}$ defined as
\begin{equation}
\mpi^{(1)\,2}=\mpii{0}^2+\Sigma_\p(\mpi^{(1)\,2})\, ,
\end{equation}
\begin{equation}
Z_\p^{(1)}
= \Big(1-\frac{d}{dq^2}\Sigma_\p(q^2)\Big|_{q^2=\mpi^{(1)\,2}}\Big)^{-1}\, .
\label{Zpi1}
\end{equation}
%
\section{Results}
%
\label{Sectresults}
\subsection{Determination of model parameters}
\label{Sectparamfit}
Our model has two groups of parameters, the bare parameters of the 
Lagrangian,
\[
\mu^2,\quad \lambda^2,\quad c,\quad \xi,\quad m_0^2\,,\quad g,\quad
h_1\,,\quad h_2\,,\quad f,
\]
and the finite parts of the counterterms,
\[
\ovd{Z}_\Phi\,,\quad \ovd{\mu^2},\quad \ovd{\lambda^2},\quad
\ovd{Z}_Y\,,\quad \ovd{m_0^2}\,,\quad \ovd{g},\quad \ovd{h}_1\,,\quad
\ovd{h}_2\,,\quad \ovd{f}\,.
\]
However, not all of these parameters are independent of each
other. Our final results do not separately depend on $f$ and
$\ovd{f}$, but only on the combination $(f+\ovd{f})$, and the
counterterms $\ovd{m_0^2}$ and $\ovd{h}_2$ always appear in the
combination $(\ovd{m_0^2}+\ovd{h}_2\sigma_0^2)$. Thus we have altogether
16 independent parameters.

In practice it is more convenient to employ \Eqs{minimumsigma0},
(\ref{masses}), (\ref{mpi012}), (\ref{grhoeff}), and (\ref{fpitree})
and to express the bare parameters $\mu^2$, $\lambda^2$, $c$, $g$,
$h_1$\,, and $h_2$ through the more physical quantities
\[
\mpi\,,\quad \ms\,,\quad \mr\,,\quad \ma\,,\quad f_\pi\,,\quad
\mathrm{and}\quad g_{\r\p\p}^\mathrm{eff}\,.
\]

For the parameter fixing one should keep in mind that our
approximation is not self consistent. In particular, unstable
particles calculated in one-loop approximation decay into tree-level
particles. Hence one should not only fit the one-loop properties to
the empirical values, but also the tree-level masses, in order to get
the proper phenomenology. For instance, the decay $\a\rightarrow
\p\p\p$ is modeled by the decay modes $\a\rightarrow\p\r$ (dominant)
and $\a\rightarrow\p\s$ (small), with stable $\r$ and $\s$ mesons. As
we shall see, for the $\p\r$ decay mode, this works rather well for
invariant masses above $\mpi+\mr$, if both, the pion and the $\r$
meson, at tree level have already their physical masses, \ie $\mpi =
\mpi^{(1)}$~=~140~MeV and $\mr = \mr^{(1)}$. As a definition for the
$\r$-meson mass $\mr^{(1)}$ we choose the position where the
$\r$-meson spectral function has its maximum, $\mr^{(1)}=767\MeV$.

Besides the pion mass we also fit the pion decay constant to its
empirical value, $f_\pi$~=~93~MeV. Again we demand that the tree-level
value is equal to the corrected one, $f_\pi = f_\pi^{(1)}$. Among
others, this will be important for later applications of the model at
non-vanishing temperatures, where the effects will be dominated by
thermally excited tree-level pions.

Similarly, we demand that $\ma$ and $\ms$ at tree-level coincide with
the corrected $\a$- and $\s$-meson masses, \ie $\ma^{(1)}=\ma$ and
$\ms^{(1)}=\ms$. The corrected masses are again defined as the
positions where the spectral functions have their maxima. As we will
discuss below, one should not blindly fit the $a_1$-meson mass to any
value given in the literature. Instead, we will use the $\a$-meson
mass to fit the peak position of the axial vector spectral function
$a_1(q^2)$. For the $\s$-meson, which is not a well established
resonance, we will take two different values, namely 600~MeV and
800~MeV.

The width of the $\r$ meson depends on two parameters,
$g_{\r\p\p}^\mathrm{eff}$ and $\ovd{Z}_Y$\,. The latter has also
influence on the $\a$ meson width. Hence, both parameters can be fixed
by fitting the widths of the vector and axial-vector spectral
functions, $v_1(q^2)$ and $a_1(q^2)$.

The width of the vector spectral function $v_1(q^2)$ being fixed, its
height can be adjusted with the parameter $(f+\ovd{f})$. The height of
the axial vector spectral function $a_1(q^2)$ is then adjusted by
changing $\ovd{g}$, which, in practice, is quite complicated, since
this parameter changes also many other quantities because of $\p-\a$
mixing.

The parameter $\xi$ acts as a cutoff for the momenta of longitudinally
polarized vector mesons. If we set arbitrarily $\xi = 0.1$, the masses
of the longitudinally polarized vector mesons are $\gtrsim 2\GeV$,
which lies still in the range of hadronic scales, but is high enough
to avoid unphysical thresholds in the energy range of interest. Since
the masses of the longitudinally polarized vector mesons behave as
$1/\sqrt{\xi}$, we can obtain reasonable fits also for smaller values
of $\xi$, \eg for $\xi = 0.025$.

The remaining undetermined parameter is $m_0$, the vector-meson mass
in the restored phase. This value is partially constrained by the
upper limit of the $\a\rightarrow\p\s$ branching ratio. The relation
between $m_0$ and the $\a\p\s$ vertex function $\Gamma_{\a\p\s}$ is
the following. $\Gamma_{\a\p\s}$ has two contributions, corresponding
to the diagrams (a) and (b) in \Fig{FigGapr}, with the outgoing
$\r$-meson lines replaced by $\s$-meson lines. The sum of both terms
yields
\begin{equation}
\big(\Gamma_{\a\p\s}^\mathrm{(a+b)}\big)^\mu_{ab} = -g\delta_{ab}\Big(
  \frac{m_0^2-\xi k^2}{\ma^2-\xi k^2} 2 k^\mu+q^\mu\Big)\, .
\end{equation}
For transversely polarized $\a$ mesons, the $q^\mu$ term does not
contribute. Thus, for $\xi\mpi^2\ll\ma^2$, the partial decay width
$\a\rightarrow\p\s$ is proportional to $m_0^4$\,. However, even for
$m_0=0$ the axial-vector spectral function $a_1(q^2)$ has small $\p\s$
contributions from the direct $W-\p-\s$ coupling (first four diagrams
in \Fig{FigSigWW}).

It turns out that we can achieve reasonable fits for $m_0\lesssim
400\MeV$. Of course a particularly interesting case is $m_0=0$, which
corresponds to a vanishing vector meson bare mass in the restored
phase (``Brown-Rho scenario''). Therefore, in the following we will
present fits with $m_0=0$ as well as with the maximum possible value
$m_0 = 400\MeV$. Together with the two values for the $\sigma$-meson
mass this leads to four different parameter sets, which are listed in
\Tab{Tabfit}. As explained in \Sect{Sectdimreg}, the scale parameter
$\Lambda$ can be chosen arbitrarily, since the dependence of the final
results on $\Lambda$ can be absorbed in the finite parts of the
counterterms. On the other hand, the numerical values of the finite
parts of the counterterms are meaningful only if the value of
$\Lambda$ is known, and this is the reason why it is listed in
\Tab{Tabfit}.
\begin{table}
\[
\newcommand{\ce}[1]{\hfill\mbox{#1}\hfill}
\begin{array}{|l|l|r|r|r|r|}
\hline
&             &\ce{Fit A}&\ce{Fit B}&\ce{Fit C}&\ce{Fit D}\\
\hline
\raisebox{0pt}[0pt][0pt]{\mbox{\parbox[t]{3cm}{\raggedright Input}}}
&\mpi\,/\MeV  & 140      & 140      & 140      & 140      \\
&\ms\,/\MeV   & 600      & 600      & 800      & 800      \\
&f_\p\,/\MeV  & 93       & 93       & 93       & 93       \\
&\xi          & 0.1      & 0.1      & 0.1      & 0.1      \\
&m_0\,/\MeV   & 0        & 400      & 0        & 400      \\
&\Lambda\,/\MeV
              & 1000     & 1000     & 1000     & 1000     \\
\hline
\raisebox{0pt}[0pt][0pt]{\mbox{\parbox[t]{3cm}{\raggedright
  Obtained from fit of $v_1$ and $a_1$ and from conditions specified in the
  text of \Sect{Sectparamfit}}}}
&\mr\,/\MeV   & 767      & 767      & 767      & 767      \\
&\ma\,/\MeV   & 1064     & 1062     & 1056     & 1054     \\
&g_{\r\p\p}^\mathrm{eff}
              & 7.63     & 7.77     & 7.92     & 7.97     \\
&(f+\ovd{f})  & 0.076    & 0.078    & 0.083    & 0.084    \\
&\ovd{Z}_\Phi & -9.36    & -9.30    & -8.48    & -8.56    \\
&\ovd{\mu^2}\,/\MeV^2
              & 6666^2   & 3873^2   & 6211^2   & 3522^2   \\
&\ovd{\lambda^2}
              & -3147    & -1421    & -2671    & -1221    \\
&\ovd{Z}_Y    & 0.260    & 0.064    & 0.069    & -0.056   \\
&(\ovd{m_0^2}+\ovd{h}_2\s_0^2)\,/\MeV^2
              & 2352^2   & 2197^2   & 2245^2   & 2128^2   \\
&\ovd{h_1}    & 20.1     & 14.9     & 15.4     & 9.11     \\
&\ovd{g}      & 33.7     & 32.5     & 29.3     & 29.0     \\
\hline
\raisebox{0pt}[0pt][0pt]{\mbox{\parbox[t]{3cm}{\raggedright Computed from
  parameters listed above}}}
&\s_0\,/\MeV  & 150.9    & 152.5    & 153.7    & 154.1    \\
&\mpii{0}\,/\MeV
              & 86.1     & 85.3     & 84.6     & 84.4     \\
&\lambda^2    & 7.74     & 7.58     & 13.4     & 13.3     \\
&c\,/\MeV^3   & 103.8^3  & 103.5^3  & 103.2    & 103.1    \\
&g            & 5.55     & 5.52     & 5.47     & 5.45     \\
&h_1          & 23.9     & 23.2     & 22.3     & 22.0     \\
&h_2          & 25.8     & 18.4     & 24.9     & 18.0     \\
&-i\overline{T}/\ms^2\,/\MeV
              & 9.80     & 14.40    & 7.65     & 10.82    \\
\hline
\end{array}
\]
\caption{\small Model parameters and related quantities in four
  different fits.}
\label{Tabfit}
\end{table}
%
\subsection{The \boldmath{$\r$} meson and the vector correlator}
\begin{figure}
\epsfig{file=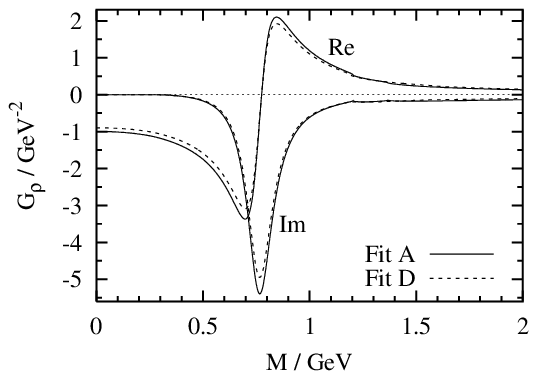,width=8cm}
\hfill
\epsfig{file=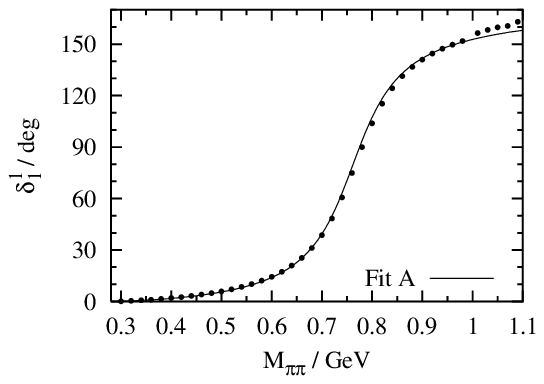,width=8cm}
\caption{\small Left panel: the Real- and imaginary part of the $\r$-meson
  propagator as a function of $M = \sqrt{q^2}$ for the parameter sets A
  and D listed in \Tab{Tabfit}. Right panel: the $\p\p$ scattering phase
  shifts in the $J=1, I=1$ channel for parameter set A (the three other
  parameter sets give the same result). The data are taken from
  \Ref{FroggattPetersen}.}
\label{FitGrhodelta11}
\end{figure}

%
The real- and imaginary parts of the $\r$-meson propagators $G_\r(q^2)
= 1/(q^2-\mr^2-\Sigma_\r^t(q^2))$ obtained for parameter sets A and D
of \Tab{Tabfit} are displayed in the left panel of
\Fig{FitGrhodelta11}. As a result of the fitting procedure described
in the previous section the propagators look very similar for both
sets. In fact, the curves for the two other parameter sets, B and C,
lie in between.

In the right panel of \Fig{FitGrhodelta11} we show the $\p\p$
scattering phase shifts in the $J=1, I=1$ channel. Near the
resonance, $s\approx \mr^2$, the $\p\p$ $T$-matrix in this channel is
clearly dominated by the $s$-channel $\rho$-meson exchange. Thus the
scattering phase shifts are given by $\tan\delta_1^1 = \Im G_\r/\Re
G_\r$. As can be seen in the figure, the phase shifts obtained in
this way fit the data up to $1\GeV$. In this region the four parameter
sets of \Tab{Tabfit} lead to practically identical results.

\begin{figure}
\epsfig{file=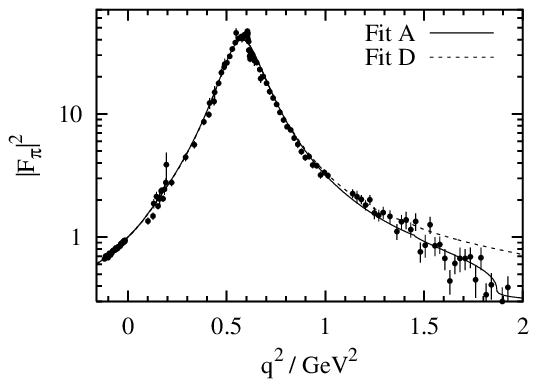,width=8cm}
\hfill
\epsfig{file=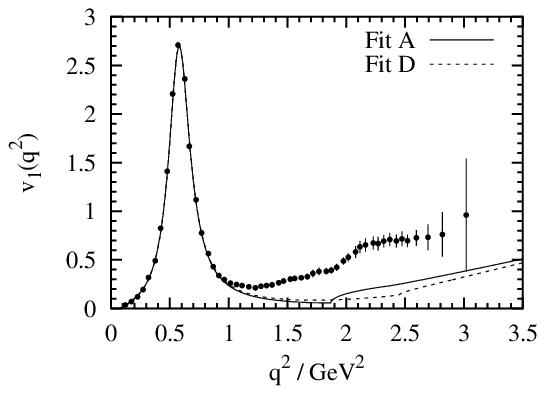,width=8cm}
\caption{\small Left panel: the electromagnetic form factor of the pion
  for the parameter sets A and D listed in \Tab{Tabfit}. The curves for
  the parameter sets B and C lie in between. The data are taken from
  \Refs{Amendolia,Barkov}. Right panel: the spectral function $v_1(q^2)$,
  together with data obtained from $\tau$ decay by the ALEPH
  collaboration \cite{ALEPH}.}
\label{FitFpiv1}
\end{figure}

The electromagnetic form factor of the pion is not simply proportional
to the $\r$-meson propagator as in Sakurai's vector dominance model,
in which the photon can couple to the two pions only via the $\r$
meson. Instead, we have to consider also the direct coupling of the
photon to two pions. Thus the pion electromagnetic form factor is
given by
\begin{equation}
F_\p(q^2) = 1+\frac{q^2 F_{\gamma\r}(q^2) g_{\r\p\p}^\mathrm{eff}}
  {q^2-\mr^2-\Sigma_\r^t(q^2)}\, .
\end{equation}
The result can be seen in the left panel of \Fig{FitFpiv1}, together with
the experimental data.
Except for the small structure at $q^2=0.61\GeV^2$ caused by
$\r-\omega$ mixing, the data are well reproduced up to $q^2\approx
1.5\GeV^2$ (fit D) or even higher (fit A).

In the right panel of \Fig{FitFpiv1} we display our results for the
vector-isovector spectral function $v_1(q^2)$. In the resonance
region, the experimental data are perfectly reproduced, but in the
continuum region, above $q^2\approx 1\GeV^2$, we get only a small
fraction of the strength measured by the ALEPH collaboration. This is
obviously related to the fact that above $1\GeV$ four-pion channels
become important. In our model, these are partially contained as
$\p\a$, $\s\r$ and $\r\r$ channels. The threshold for the $\s\r$
channel at $q^2=1.87\GeV^2$ for fit A and $q^2=2.46\GeV^2$ for fit D
is clearly visible, and also the $\r\r$ channel gives a sizable
contribution above $2.35\GeV^2$, whereas the $\p\a$ channel
contributes almost nothing. At first sight this is surprising, since
in the $\r$-meson spectral function the $\p\a$ channel is relatively
strong (in the left panel of \Fig{FitGrhodelta11} one can see the
threshold as a small kink at $1.2\GeV$). However, in the vector
spectral function two amplitudes interfere destructively, namely
$(\gamma\rightarrow\p\a)$ and
$(\gamma\rightarrow\r\rightarrow\p\a)$. This interference is absent in
the $\s\r$ and $\r\r$ channels, because the $\s\r$ channel can be
reached only via the $\r$ meson
$(\gamma\rightarrow\rho\rightarrow\s\r)$, while for the $\r\r$ channel
only the direct coupling exists $(\gamma\rightarrow\r\r)$.
%
\subsection{The \boldmath{$\a$} meson and the axial-vector correlator}
%

Before the precise $\tau$ decay measurements by the ALEPH
collaboration \cite{ALEPH}, the knowledge about the axial vector
correlator and the properties of the $a_1$ meson was more or less
uncertain, and even in the latest particle data book \cite{PDB}, mass
and width of the $a_1$ meson still have large uncertainties: $\ma =
1230\pm 40\MeV$ and $\Gamma_\a = 250$ to $600\MeV$. However, as
pointed out in \Ref{Isgur}, the discrepancy of different experiments
to some extent stems from the use of different parametrizations of the
$\a$ propagator in the analyses. Therefore, one must be careful when
taking over parameters from another model. We have decided to fit the
axial-vector spectral function ourselves and not to restrict the
parameters to the values listed in the particle data book.

\begin{figure}
\epsfig{file=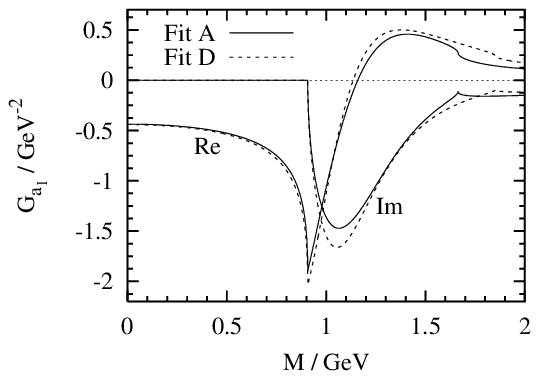,width=8cm}
\hfill
\epsfig{file=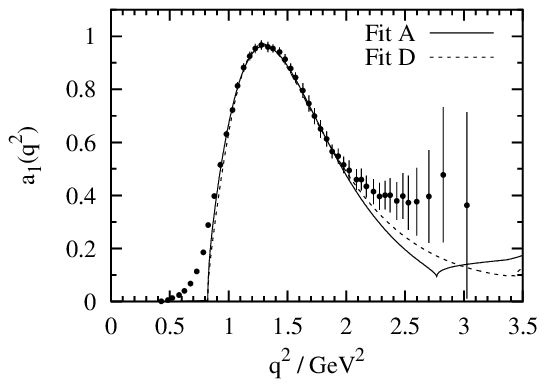,width=8cm}
\caption{\small Left panel: the real- and imaginary part of the $\a$
  propagator as a function of $M = \sqrt{q^2}$ for the parameter sets A
  and D listed in \Tab{Tabfit}. The curves for the parameter sets B and
  C lie in between. Right panel: the transverse axial-vector spectral
  function $a_1$ as a function of $q^2$. The data are taken from
  \Ref{ALEPH}.}
\label{FitGa1a1}
\end{figure}

The resulting $\a$ propagator $G_\a(q^2) =
1/(q^2-\ma^2-\Sigma_\a^t(q^2))$ is shown in the left panel of
\Fig{FitGa1a1}.
The mass, defined as the position of the maximum of $-\Im G_\a$, is
rather low (see \Tab{Tabfit}). However, if we adopt another usual
convention, defining the mass as the position of the zero of $\Re
G_\a$, we obtain higher values between $\ma=1136\MeV$ (fit D) and
$\ma=1159\MeV$ (fit A). The $\a$ meson width is defined as the
difference of the two masses where $-\Im G_\a$ crosses one half of its
maximum. We find $\Gamma_\a=424\MeV$ for fit A and $\Gamma_\a=393\MeV$
for fit D.

The spectral function of the transverse axial-vector correlator,
$a_1(q^2)$, is displayed in the right panel of \Fig{FitGa1a1}. Between
$0.9$ and $2\GeV^2$ we can fit the ALEPH data very well. In this range
the spectral function is almost saturated by three-pion states
\cite{ALEPH}, which in our model are approximated by $\p\r$ states
with sharp $\r$ mesons. Over a wide range of $q^2$ this seems to be
adequate. Only in the region $q^2\lesssim (\mpi+\mr)^2$, of course,
some strength is missed. Above $2\GeV^2$ the spectral function
receives sizeable contributions from five-pion states
\cite{ALEPH}. These are partially contained in our model as $\s\a$ and
$\r\a$ states (for fit A, the $\s\a$ threshold is visible at
$2.77\GeV^2$), but this is obviously not sufficient in order to fit
the data.

\begin{figure}
\epsfig{file=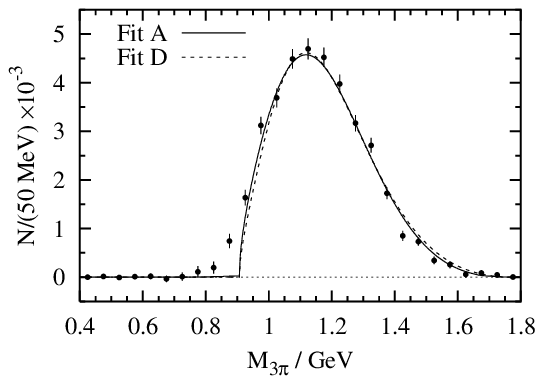,width=8cm}
\hfill
\epsfig{file=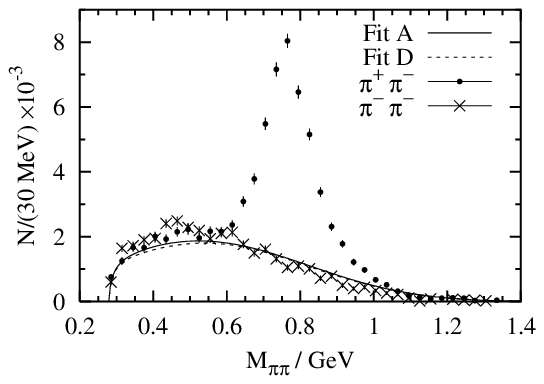,width=8cm}
\caption{\small Left panel: the three-pion invariant mass distribution of
  $\tau^- \rightarrow \p^-\p^-\p^+\nu_\tau$ for the parameter sets A and
  D listed in \Tab{Tabfit}. Right panel: the invariant mass distributions of
  $\p^-\p^-$ (theoretical curves and crosses) and $\pi^+\pi^-$ pairs
  (dots) of $\tau^- \rightarrow \p^-\p^-\p^+\nu_\tau$\,. The data are
  taken from \Refs{ARGUS} and \cite{Walther}.}
\label{FitARGUS}
\end{figure}

In \Fig{FitARGUS} it is shown that up to the $\tau$ mass the exclusive
three-pion final state can very well be approximated by a $\p\r$ final
state with a sharp $\r$ meson.
In the left panel we display the three-pion invariant mass
distribution measured by the ARGUS collaboration \cite{ARGUS} and
obtained in our model as the sum of $1/2$ of the $\p\r$ and $2/3$ of
the $\p\s$ spectrum (see \App{Appm3pi}). 
The $\p\s$ channel contributes
only between $1.3\,\%$ (fit C) and $5.4\,\%$ (fit B) to the integrated
spectrum, which lies below the upper limit of $6\,\%$ given in
\Ref{ARGUS}.

In the right panel of \Fig{FitARGUS} we display the $\p^-\p^-$ and
$\p^+\p^-$ invariant mass distributions of the decay $\tau^-
\rightarrow \p^-\p^-\p^+\nu_\tau$. The experimental data show that the
$\p^+\p^-$ distribution (dots) is the sum of the $\p^-\p^-$
distribution (crosses) and a distribution which is concentrated around
the $\r$-meson mass, indicating that the $\p^-\p^-\p^+$ final state is
indeed reached via a $\p^-\r^0$ intermediate state. The interference
term, which is responsible for the fact that at low invariant masses
the $\p^+\p^-$ distribution lies below the $\p^-\p^-$ distribution, is
small. The $\p^-\p^-$ distribution which corresponds to our model (see
\App{Appm2pi}) is in good agreement with the data. Since we do not take
into account the $\rho$-meson width and the interference term, the
corresponding $\p^+\p^-$ distribution would be given by the $\p^-\p^-$
distribution plus a delta function at the $\r$-meson mass (and a small
delta function at the $\s$-meson mass).
%
\subsection{The longitudinal axial-vector correlator}
\begin{figure}
\begin{center}
\epsfig{file=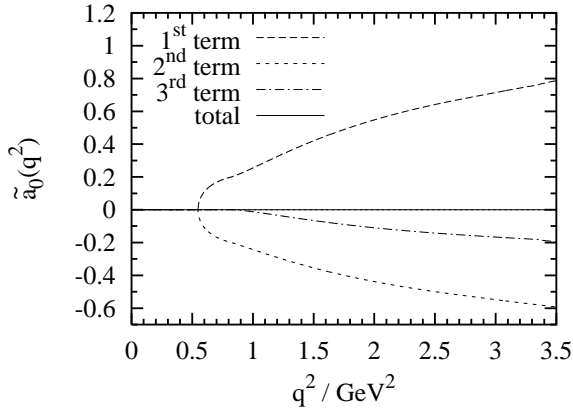,width=8cm}
\end{center}
\vspace{-0.5cm}
\caption{\small The multi-particle contribution of the longitudinal
  axial-vector spectral function, $\tilde{a}_0(q^2)$, calculated with
  the parameters from fit A in \Tab{Tabfit}. For the meaning of the
  different terms see \Eq{SigWl}.}
\label{Fita0}
\end{figure}

%
As one can see from \Eqs{SigWl} to (\ref{Zpi1}) the longitudinal
axial-vector correlator has a pole at $q^2 = \mpi^{(1)\,2}\ (= \mpi^2)$
with residue $4 \pi^2 f_\p^{(1)\,2}\ (= 4 \pi^2 f_\p^2)$, \ie
\begin{equation}
a_0(q^2) = 4\pi^2 f_\p^2 \delta(q^2-\mpi^2) + \tilde{a}_0(q^2)\, ,
\label{a0tilde}
\end{equation}
where $\tilde{a}_0(q^2)$ denotes the contribution of multi-particle
channels. According to \Eq{SigWl}, the longitudinal $W$-boson self
energy consists of three terms. The corresponding contributions to
$\tilde{a}_0(q^2)$ are shown in \Fig{Fita0}.
The first term gives a positive contribution to $\tilde{a}_0(q^2)$,
whereas the other terms contain interferences of several amplitudes
and are negative. The sum of the three terms is of course positive,
but almost zero. In fact, it is not visible in the figure since it is
about a factor $10^{-4}$ smaller than the first contribution. To be
more quantitative, we integrate the strength of $\tilde{a}_0(q^2)$ up
to $m_\tau=1777\MeV$. With the parameters of fit A we find
$\int_0^{m_\tau^2} dq^2\,\tilde{a}_0(q^2)=75\MeV^2$, which has to be
compared with the strength $4\pi^2 f_\pi^2 = 0.34\GeV^2$ of the
single-pion contribution.

The smallness of $\tilde{a}_0(q^2)$ is a manifestation of the PCAC
hypothesis. Usually the PCAC hypothesis is formulated as
$\partial^\mu \vec{j}_\mu^5 = f_\p \mpi^2 \vec{\p}$\,. However, beyond
tree level this is almost meaningless, since the field operator $\p_a$
has non-vanishing matrix elements $\langle f|\p_a|0\rangle$ not only
for single-pion states $|f\rangle$, but also for multi-particle
states, \eg in our model for $\p\s$, $\p\r$, $\a\s$ and $\a\r$
states. In fact, as pointed out by Weinberg \cite{Weinberg2}, the PCAC
assumption should be interpreted in a different way, namely that
$\tilde{a}_0(q^2)$, which is directly related to the multi-particle
content of $\partial^\mu j_{\mu\,a}^5|0\rangle$, is negligible.
Obviously this is fulfilled very well in our model.
%
\subsection{The \boldmath{$\sigma$}-meson propagator}
\begin{figure}
\epsfig{file=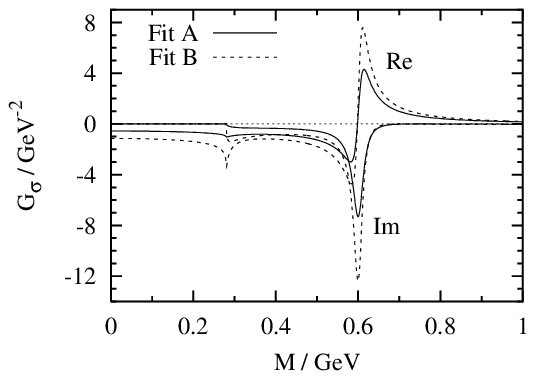,width=8cm}
\hfill
\epsfig{file=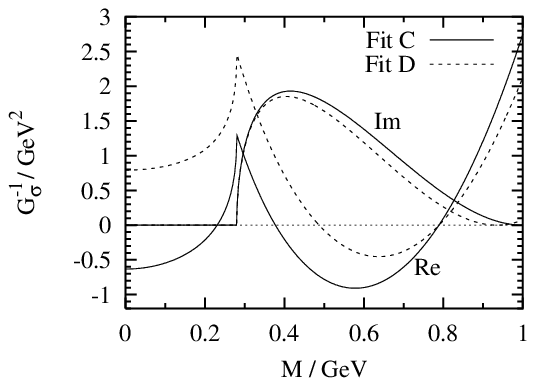,width=8cm}
\caption{\small Left panel: the real- and imaginary part of the $\s$-meson
  propagator for parameter sets A and B. Right panel: the real- and imaginary
  parts of the inverse $\s$-meson propagator for parameter sets C and D.}
\label{FitGsigma}
\end{figure}

%
A general problem of models with linearly realized chiral symmetry is
that at tree level it predicts the existence of a sharp scalar
isoscalar meson, which has not been found experimentally. However, in
the latest particle data book \cite{PDB} a meson with these quantum
numbers is listed as ``$f_0(400-1200)$ or $\s$''. This particle has a
very large width (approximately as large as the mass) due to the decay
into two pions. Since our calculation contains the $\s\rightarrow\p\p$
graph, one could expect that it should be possible to get a
satisfactory description of the $\s$ meson. Unfortunately, this is not
the case for the following reasons:

The $\s\p\p$ vertex has four contributions, analogous to the four
contributions to the $\r\p\p$ vertex of the gauged linear $\s$ model
shown in \Fig{FigGrpp}. These amplitudes interfere destructively. If
the outgoing pions are on the mass shell, the total vertex function
has a zero at some value of the invariant mass $M$ of the $\s$
meson. For the parameter sets A and B listed in \Tab{Tabfit}, this
zero lies at $754\MeV$ and $705\MeV$, respectively. This is above the
tree-level $\s$ mass, $m_\s = 600\MeV$, but the amplitude is already
very small for $M\approx\ms$. The authors of \Ref{KoRudaz} hoped that
this effect could improve the description of the $\s$ meson, since
they believed that the width obtained from the linear $\s$ model
without vector mesons was too large. However, now the width is much
too small, as can be seen in the left panel of \Fig{FitGsigma}, where
we display the $\s$-meson propagator for the parameter sets A and
B. The different normalization results from the fact that we adjust
the wave-function renormalization constant $\ovd{Z}_\Phi$ for the pion
propagator and not for the $\s$-meson propagator.

Another problem arises from the fact that, at the two-pion threshold,
the $\s$-meson self energy usually becomes strongly attractive
(\ie negative). Since the $\s\p\p$ vertex function at low $q^2$ is
dominated by the ordinary $\s$ model vertex $\propto\lambda^2\s_0$,
with $\lambda^2 \propto \ms^2-\mpii{0}^2$, this effect is stronger for
higher $\s$-meson masses and has therefore more dramatic consequences
for the parameter sets C and D. For parameter set C we find a two-pion
bound state at $M = 231\MeV$, while for parameter set D the $\s$-meson
propagator has the wrong sign at low $M$ and a pole in the complex
plane. For illustration, we have plotted the inverse $\s$-meson
propagator in the right panel of \Fig{FitGsigma}.

Obviously the $\s$-meson cannot be described in one-loop
approximation. A good description of the $\s$-meson in the linear $\s$
model can be obtained by summing the $\p\p$ loops in the self energy
to all orders (RPA) in the large-$N$ limit ($N$ = number of pions)
\cite{Aouissat2}. Summing the $\p\p$ loops reduces the strong
attraction in the real part at the two-pion threshold. However, the
application of the RPA to our Lagrangian would be extremely
difficult. On the other hand, for the vector mesons, in which we are
interested, the one-loop calculation seems to be sufficient.
%
%
\section{Conclusions}
%
\label{Sectconclusions}
We have presented a chirally symmetric hadronic model for the vector
and axial-vector correlators in vacuum. The starting point was the
linear $\sigma$ model extended by $\rho$-meson and $a_1$ fields in a
chirally symmetric way. Unlike the gauged linear $\sigma$ model,
which leads to phenomenological difficulties (see \Sect{SectLgauged}),
our Lagrangian is only symmetric under global, but not under local
$SU(2)_L\times SU(2)_R$ transformations. After all, chiral symmetry is
also only an (approximate) global symmetry in QCD.

For a realistic description of $\rho$ and $a_1$ we could not stay at
tree-level, where all mesons are stable particles. In
order to include the most important decay channels,
$\rho\rightarrow\pi\pi$ and $a_1\rightarrow\rho\pi$, we performed a
one-loop calculation. Here care had to be exercised, not to destroy
the symmetries of the model by choosing an unsuited regularization
scheme. We have decided to use dimensional regularization in order to
isolate the divergences, which are then subtracted by chirally
symmetric counterterms.

The price to pay for giving up the requirement of a local chiral
symmetry is the loss of the so-called current-field identity which
directly relates the vector and axial-vector currents to the meson
fields. Therefore some additional work was needed to construct the
correlators in a consistent way.

In the last part of the paper we have presented numerical results. We
have fixed our parameters to reproduce the empirical values of $m_\pi$
and $f_\pi$ and the peak regions of the vector and axial vector
spectral functions, which have been extracted by the ALEPH
collaboration from the $\tau$ decay. Basically, this leaves two
parameters undetermined, namely the mass of the $\sigma$-meson, which
is not a well established resonance in nature, and the mass parameter
$m_0$. The latter is the fraction of the bare $\rho$-meson mass which
is not proportional to the vacuum expectation value of the $\sigma$
field, \ie the fraction which would survive in the restored
phase. From our fit we only found an upper limit of $400\MeV$ for
$m_0$. On the other hand, a particularly interesting case is $m_0=0$,
which corresponds to a vanishing bare vector meson mass in the
restored phase (``Brown-Rho scenario''). Therefore, we have performed
calculations with $m_0=0$ and with $m_0 = 400\MeV$, both for two
different $\s$-meson masses, $\ms = 600\MeV$ and $\ms = 800\MeV$.

All fits describe the peak regions of the experimental spectral
functions equally well and also give a satisfactory description of the
pion electromagnetic form factor, the $\pi\pi$ phase shifts in the
vector-isovector channel, and the $\p^-\p^-\p^+$ and $\p^-\p^-$ mass
distributions for the decay $\tau^- \rightarrow
\p^-\p^-\p^+\nu_\tau$\,. At higher energies, $q^2 \geq 1\GeV^2$ in the
vector channel and $q^2 \geq 2\GeV^2$ in the transverse axial-vector
channel our fits underestimate the empirical spectral functions. In
these regions four-pion (vector channel) or five-pion (axial-vector
channel) final states become relevant, which are not properly
contained in our one-loop calculation.

The longitudinal axial correlator is completely dominated by the pion
pole, in agreement with the PCAC hypothesis. The remaining strength,
which is due to the small explicit breaking of chiral symmetry in our
Lagrangian, is the result of a delicate cancellation of three terms
which are four orders of magnitude larger than their sum. This nicely
demonstrates the consistency of our model with chiral symmetry as well
as the stability of the numerics.

In the present paper we have restricted ourselves to vacuum
properties. However, having the technology developed, it is now
straightforward to calculate the correlators at finite
temperatures. Work in this direction is in progress. The extension to
finite baryon density in a chirally symmetric way is obviously more
difficult.
\vspace{1cm}
\begin{center}
\large \bf Acknowledgements
\end{center}
One of us (M.\@ U.) acknowledges support from the Studienstiftung des
deutschen Volkes. This work was supported in part by the BMBF.
\vspace{1cm}
\begin{appendix}
%
\section{The three-pion invariant mass distribution of
\boldmath{$\tau^-\rightarrow\p^-\p^-\p^+\nu_\tau$}}
%
\label{Appm3pi}
The differential $\tau$ decay width has the form
\begin{equation}
\frac{d\Gamma_{\tau\rightarrow \p^-\p^-\p^+\nu_\tau}}{dM_{3\p}} = 
  G_F^2\cos^2{\theta_C}\frac{(m_\tau^2-M_{3\p}^2)^2}{2\pi^2 m_\tau^3 M_{3\p}}
  \big(m_\tau^2 H_{\p^-\p^-\p^+}^l
  +(m_\tau^2+2 M_{3\p}^2) H_{\p^-\p^-\p^+}^t\big)\, .
\end{equation}
In this formula, $G_F = e^2/(4\sqrt{2}\sin^2\theta_W m_W^2)$ denotes
Fermi's coupling constant and $H_{\p^-\p^-\p^+}^l$ and
$H_{\p^-\p^-\p^+}^t$ are the longitudinal and transverse components of
the so-called hadron tensor, respectively. Under the assumption of
PCAC, the longitudinal component can be neglected.

In our model the $\p^-\p^-\p^+$ spectrum consists of $1/2$ of the
$\p\r$ plus $2/3$ of the $\p\s$ spectrum, \ie
\begin{equation}
H_{\p^-\p^-\p^+}^t(M_{3\p}^2) = 
  \frac{1}{2} H_{\p\r}^t(M_{3\p}^2)+\frac{2}{3} H_{\p\s}^t(M_{3\p}^2)\, .
\end{equation}
The $\p\r$ and $\p\s$ hadron tensors are given by
\begin{align}
H_{\p\r}^t(q^2) =&\frac{1}{3}
  \Big(\frac{q_\mu q_{\mu^\prime}}{q^2}-g_{\mu\mu^\prime}\Big)
  \intd{3}{k_\p}\frac{1}{2 k_\p^0} \intd{3}{k_\r}\frac{1}{2 k_\r^0}
  (2\pi)^4 \delta(q-k_\p-k_\r) \nonumber\\
& \times \zp{1}
  \Big(\frac{k_{\r\,\nu}k_{\r\,\nu^\prime}}{\mr^2}-g_{\nu\nu^\prime}\Big)
  \big(A_{W\p\r}\big)_{1bc}^{\mu\nu}
  \big(A_{W\p\r}^{*}\big)_{1bc}^{\mu^\prime\nu^\prime}\, ,\\
H_{\p\s}^t(q^2) =&\frac{1}{3}
  \Big(\frac{q_\mu q_{\mu^\prime}}{q^2}-g_{\mu\mu^\prime}\Big)
  \intd{3}{k_\p}\frac{1}{2 k_\p^0} \intd{3}{k_\s}\frac{1}{2 k_\s^0}
  (2\pi)^4 \delta(q-k_\p-k_\s) \nonumber\\
& \times \zp{1}
  \big(A_{W\p\s}\big)_{1b}^\mu\big(A_{W\p\s}^{*}\big)_{1b}^{\mu^\prime}\, ,
\end{align}
with $k_\p^0 = \sqrt{\mpi^2+\vec{k}_\p^2}$\,,
$k_\r^0 = \sqrt{\mr^2+\vec{k}_\r^2}$\,,
$k_\s^0 = \sqrt{\ms^2+\vec{k}_\s^2}$ and
\begin{align}
\big(A_{W\p\r}\big)_{abc}^{\mu\nu} =&
  \frac{\sin\theta_W}{e\cos{\theta_C}}\Big(
  \big(\Gamma_{W\p\r}\big)_{abc}^{\mu\nu}
  +\big(\Gamma_{W\a}\big)_{ad}^{\mu\kappa}
  i\big(G_\a(q^2)\big)_{\kappa\lambda}
  \big(\Gamma_{\a\p\r}\big)_{dbc}^{\lambda\nu}\Big)\, ,\\
\big(A_{W\p\s}\big)_{ab}^\mu =&
  \frac{\sin\theta_W}{e\cos{\theta_C}}\Big(
  \big(\Gamma_{W\p\s}\big)_{ab}^\mu
  +\big(\Gamma_{W\a}\big)_{ad}^{\mu\kappa}
  i\big(G_\a(q^2)\big)_{\kappa\lambda}
  \big(\Gamma_{\a\p\s}\big)_{db}^\lambda\Big)\, .
\end{align}
The vertex functions $\Gamma_{W\p\r}$ and $\Gamma_{W\p\s}$ are defined
analogously to \Fig{FigGapr}.

The curves shown in the left panel of \Fig{FitARGUS} are normalized as
follows:
\begin{equation}
\frac{dN_{\p^-\p^-\p^+}}{dM_{3\p}}
= N_{\tau^-}\tau_{\tau^-}
  \frac{d\Gamma_{\tau^-\rightarrow\p^-\p^-\p^+\nu_\tau}}{dM_{3\p}}
= \frac{N_{\p^-\p^-\p^+}\tau_{\tau^-}}
  {B(\tau^-\rightarrow\p^-\p^-\p^+\nu_\tau)}\,
  \frac{d\Gamma_{\tau^-\rightarrow\p^-\p^-\p^+\nu_\tau}}{dM_{3\p}}\, ,
\end{equation}
where $N_{\p^-\p^-\p^+} = 37170$ is the total number of $\p^-\p^-\p^+$
events \cite{ARGUS}, $\tau_{\tau^-} = 290.0\,\mathrm{fs}$ is the
$\tau$ lifetime \cite{ALEPH}, and
$B(\tau^-\rightarrow\p^-\p^-\p^+\nu_\tau) = 9.15\,\%$ is the branching
ratio \cite{ALEPH}.
%
\section{The \boldmath{$\p^-\p^-$} invariant mass distribution of
\boldmath{$\tau^-\rightarrow\p^-\p^-\p^+\nu_\tau$}}
%
\label{Appm2pi}
Since our model does not contain the $\p^-\p^-\p^+$ final state, but
only $\pi^-\rho^0$ and $\p^-\sigma$ final states, the $\p^-\p^-$
distribution cannot be computed directly. Instead we must interprete
the $\r^0$ or $\s$ meson in the final state, with momentum $k_\r$ or
$k_\s$\,, respectively, as a $\p^+\p^-$ pair with momenta $p_{\p^+}$
and $p_{\p^-}$ (\ie $p_{\p^+}+p_{\p^-} = k_\r$ or $k_\s$\,,
respectively). For simplicity let us first look at the $\p\s$ final
state.

In the rest frame of the $\s$ meson the kinematics is given by
\begin{align}
k_\s &= (\ms,\vec{0})\,,
\nonumber\\
k_{\p^-} & = (q_0-\ms,\vec{q})\,,\quad
  q_0 = \frac{q^2-\ms^2-\mpi^2}{2\ms}\,,\quad
  |\vec{q}| = \frac{\sqrt{(q^2-\mpi^2-\ms^2)^2-4\mpi^2\ms^2}}{2\ms}\,,
\nonumber\\
p_{\p^{\pm}} & = \Big(\frac{\ms}{2}, \mp\vec{p}\Big)\,,\quad
  |\vec{p}| = \sqrt{\frac{\ms^2}{4}-\mpi^2}\,,\quad
  \vec{p}\cdot\vec{q} = |\vec{p}| |\vec{q}| z\,.
\end{align}
Thus the $\p^-\p^-$ invariant mass depends only on $q^2$ and $z =
\cos(\vec{p},\vec{q})$:
\begin{equation}
M_{2\p}^2 = \frac{q^2-\ms^2+3\mpi^2}{2}-2|\vec{p}| |\vec{q}| z\,.
\end{equation}
Let $P_\s(z)$ be the probability for a certain angle between $\vec{p}$
and $\vec{q}$. Since the $\s$ meson decays into a $\p^+\p^-$ pair in
$s$ wave, the angular distribution in its rest frame is isotropic,
\ie $P_\s(z) = \mbox{const.} = 1/2$. Now the $\p\s$ contribution to the
$\p^-\p^-$ invariant mass spectrum can be written as
\begin{align}
\frac{dN_{\p^-\p^-}}{dM_{2\p}^2} &=
  \int_{9\mpi^2}^{\infty}\!\! dq^2\,
  \Big(\frac{dN_{\p^-\p^-\p^+}^{(\p\s)}}{dM_{3\p}^2}\Big)_{M_{3\p}^2=q^2}
\nonumber\\
& \phantom{=\int_{9\mpi^2}^{\infty}\!\! dq^2\,}\times
  \int_{-1}^1\!\! dz\, P_\s(z)
  \delta\big(M_{2\p}^2-(q^2-\ms^2+3\mpi^2)/2+2|\vec{p}| |\vec{q}| z\big)
\nonumber\\
&= \int_{q^2_-}^{q^2_+}\!\! dq^2\,
  \Big(\frac{dN_{\p^-\p^-\p^+}^{(\p\s)}}{dM_{3\p}^2}\Big)_{M_{3\p}^2=q^2}\,\,
  \frac{1}{2|\vec{p}||\vec{q}|}
  P_\s\big((q^2-\ms^2+3\mpi^2-2M_{2\p}^2)/(4|\vec{p}||\vec{q}|)\big)\,,
\end{align}
where $dN_{\p^-\p^-\p^+}^{(\p\s)}/dM_{3\p}$ is the $\p\s$ contribution
to the three-pion spectrum discussed in \App{Appm3pi}, and the
integration boundaries $q^2_{\pm}$ in the last line are given by
\begin{equation}
q^2_\pm = \frac{2 \mpi^4+\ms^2 M_{2\p}^2\pm 2|\vec{p}| \ms M_{2\p}
  \sqrt{M_{2\p}^2-4\mpi^2}}{2\mpi^2}\,.
\end{equation}

For the $\p\r$ contribution the situation is very similar if one replaces
everywhere $\ms$ by $\mr$. The only complication comes from the fact that the
angular distribution cannot be assumed to be isotropic, since the $\r$ meson
decays into a $\p^+\p^-$ pair in $p$ wave. In order to obtain the angular
distribution, we assume that the $\r$ meson is measured by a fictitious
detector, which forces it onto its mass shell without changing its spin, and
the decay into two pions takes place after this measurement. Under these
conditions the angular distribution is given by the square of the amplitude
\begin{equation}
B^\mu \varepsilon_{abc} = (A_{W\p\r})^{\mu\nu}_{abc}
  (p_{\p^-\,\nu}-p_{\p^+\,\nu})\,,
\end{equation}
averaged over the three transverse $W$ polarizations, and evaluated for the
kinematics given above (but with $\ms$ replaced by $\mr$):
\begin{equation}
P_\r(z) \propto \Big(\frac{q_\mu q_{\mu^\prime}}{q^2}-g_{\mu\mu^\prime}\Big)
  B^\mu B^{*\,\mu^\prime} = c_0 + c_2 z^2\,.
\end{equation}
The normalized angular distribution is then given by
\begin{equation}
P_\r(z) = \frac{c_0 + c_2 z^2}{2(c_0 + c_2/3)}\,.
\end{equation}
\end{appendix}
%
%

%
\end{document}